\documentclass[12pt]{article}
\pdfoutput=1 
\usepackage{amssymb}
\usepackage{amsmath}
\usepackage{amstext}
\usepackage{graphicx,epsfig}
\usepackage{epsfig}
\usepackage{verbatim} 
\usepackage{fancybox}
\usepackage{color}
\usepackage{ulem}
\usepackage{enumitem}
\usepackage{subfigure}
\usepackage{bbm}
\usepackage{parskip}
\usepackage[numbers,sort&compress]{natbib}
\usepackage{ytableau}

\usepackage{tikz}
\usetikzlibrary{decorations}
\pgfdeclaredecoration{complete sines}{initial}
{
    \state{initial}[
        width=+0pt,
        next state=upsine,
        persistent precomputation={\pgfmathsetmacro\matchinglength{
            \pgfdecoratedinputsegmentlength / int(\pgfdecoratedinputsegmentlength/\pgfdecorationsegmentlength)}
            \setlength{\pgfdecorationsegmentlength}{\matchinglength pt}
        }] {}
    \state{upsine}[width=\pgfdecorationsegmentlength,next state=downsine]{
        \pgfpathsine{\pgfpoint{0.25\pgfdecorationsegmentlength}{0.5\pgfdecorationsegmentamplitude}}
        \pgfpathcosine{\pgfpoint{0.25\pgfdecorationsegmentlength}{-0.5\pgfdecorationsegmentamplitude}}
    }
    \state{downsine}[width=\pgfdecorationsegmentlength,next state=upsine]{
        \pgfpathsine{\pgfpoint{0.25\pgfdecorationsegmentlength}{-0.5\pgfdecorationsegmentamplitude}}
        \pgfpathcosine{\pgfpoint{0.25\pgfdecorationsegmentlength}{0.5\pgfdecorationsegmentamplitude}}
}
    \state{final}{}
}

\newcommand{\Comment}[1]{{}}
\definecolor{darkblue}{rgb}{0.15,0.35,0.55}
\definecolor{reddish}{rgb}{0.65, 0.2, 0.2}
\usepackage[linktocpage=true]{hyperref}
\hypersetup{
colorlinks=true,
citecolor=darkblue,
linkcolor=reddish,
urlcolor=darkblue,
pdfauthor={},
pdftitle={},
pdfsubject={}
}

\newcommand{\be}{\begin{equation}}
\newcommand{\ee}{\end{equation}}
\newcommand{\bea}{\begin{eqnarray}}
\newcommand{\eea}{\end{eqnarray}}
\newcommand{\beas}{\begin{eqnarray*}}
\newcommand{\eeas}{\end{eqnarray*}}

\def\({\left(}
\def\){\right)}

\newcommand{\rd}{{\rm d}}

\def\gsim{ \lower .75ex \hbox{$\sim$} \llap{\raise .27ex \hbox{$>$}} }
\def\lsim{ \lower .75ex \hbox{$\sim$} \llap{\raise .27ex \hbox{$<$}} }

\def\xyma{\xymatrix@M.7em}
\def\xymas{\xymatrix@M.1em}
\newcommand{\ba}{\begin{eqnarray}}
\newcommand{\ea}{\end{eqnarray}}

\title{}
\author{}

\numberwithin{equation}{section}

\begin{document}
%
\renewcommand{\thefootnote}{\fnsymbol{footnote}}
~
\vspace{1.75truecm}
\begin{center}
{\LARGE \bf{Massive and Massless Spin-2 Scattering and}}\\ \vspace{.2cm}
{\LARGE \bf{Asymptotic Superluminality}}
\end{center} 

\vspace{1truecm}
\thispagestyle{empty}
\centerline{{\large James Bonifacio,${}^{\rm a,}$\footnote{\href{mailto:james.bonifacio@case.edu}{\texttt{james.bonifacio@case.edu}}} Kurt Hinterbichler,${}^{\rm a,}$\footnote{\href{mailto:kurt.hinterbichler@case.edu}{\texttt{kurt.hinterbichler@case.edu}}} Austin Joyce,${}^{\rm b,}$\footnote{\href{mailto:austin.joyce@columbia.edu}{\texttt{austin.joyce@columbia.edu}}} and Rachel A. Rosen${}^{\rm b,}$\footnote{\href{mailto:rar2172@columbia.edu}{\texttt{rar2172@columbia.edu}}}}}
\vspace{.5cm}
 
\centerline{{\it ${}^{\rm a}$CERCA, Department of Physics,}}
 \centerline{{\it Case Western Reserve University, 10900 Euclid Ave, Cleveland, OH 44106}} 
 \vspace{.25cm}
 
 \centerline{{\it ${}^{\rm b}$Center for Theoretical Physics, Department of Physics,}}
 \centerline{{\it Columbia University, New York, NY 10027}} 
 \vspace{.25cm}

 \vspace{.8cm}
\begin{abstract}
\noindent
We constrain theories of a massive spin-2 particle coupled to a massless spin-2 particle by demanding the absence of a time advance in eikonal scattering.  This is an $S$-matrix consideration that leads to model-independent constraints on the cubic vertices present in the theory.  Of the possible cubic vertices for the two spin-2 particles, the requirement of subluminality leaves a particular linear combination of cubic vertices of the Einstein--Hilbert type.  Either the cubic vertices must appear in this combination or new physics must enter at a scale parametrically the same as the mass of the massive spin-2 field, modulo some standard caveats. These conclusions imply that there is a one-parameter family of ghost-free bimetric theories of gravity that are consistent with subluminal scattering. When both particles couple to additional matter, subluminality places additional constraints on the matter couplings. We additionally reproduce these constraints by considering classical scattering off of a shockwave background in the ghost-free bimetric theory.

\end{abstract}

\newpage

\setcounter{tocdepth}{2}
\tableofcontents
\newpage
\renewcommand*{\thefootnote}{\arabic{footnote}}
\setcounter{footnote}{0}

\section{Introduction}

In this paper we study infrared (IR) constraints on theories containing both a massive spin-2 particle and a massless spin-2 particle in the low energy spectrum, extending the analyses of~\cite{Camanho:2014apa} and~\cite{Camanho:2016opx,Hinterbichler:2017qyt}.  Consistent effective field theories (EFTs) describing a massive spin-2 particle coupled to a massless spin-2 particle with a cutoff scale parametrically larger than its mass can be constructed by introducing dynamics for the reference metric in the EFT of a single massive spin-2 particle~\cite{ArkaniHamed:2002sp,Creminelli:2005qk}.  The theory of this type with the highest known cutoff and largest known regime of validity is ghost-free bi-gravity \cite{Hassan:2011zd}, obtained by making the reference metric of ghost-free \cite{Hassan:2011hr} massive gravity \cite{deRham:2010ik,deRham:2010kj} dynamical.  However, just as in the theory of a single massive spin-2 particle, it is not known whether any of these theories can arise as the IR effective theory of some ultraviolet (UV)-complete description.  Our goal is to find constraints that may point us in the right direction.\footnote{Similar constraints can also be placed on some theories of massless higher spins~\cite{Hinterbichler:2017qcl}.}

Absence of superluminality is a traditional constraint placed on low-energy theories.  These constraints are usually derived by looking for a classical background solution to the effective theory, and then demanding subluminal propagation of fluctuations about this solution. (This was mentioned in the context of bi-gravity in \cite{Deser:2013gpa}.)  However, it is more desirable to have a sharp $S$-matrix diagnostic of superluminality that applies asymptotically and sidesteps questions about whether the background can actually be reached dynamically within the regime of the effective theory~\cite{Burrage:2011cr,Hassan:2017ugh} or whether the superluminality is itself detectable within the effective theory~\cite{Goon:2016une}.

Eikonal scattering is a kinematic regime of $2\to2$ scattering where the center-of-mass energy squared, $s$, is taken to be large with the impact parameter held fixed.  We display the precise eikonal kinematics for particles of any mass in Section~\ref{sec:scalarspin2}.  The leading contribution to scattering in this kinematic regime is captured by a sum of ladder and crossed ladder diagrams in the $t$-channel, which exponentiate into the expression~\cite{Cheng:1969eh,Levy:1969cr,Abarbanel:1969ek}
\be
i{\cal M}_{\rm eikonal}(s, t) = 2s\int\rd^{2}\vec b\,e^{i\vec q\cdot\vec b}\left(e^{i\delta(s,\vec b)}-1\right)\,.
\ee
Here $\vec b$ is the impact parameter, Fourier conjugate to the exchanged momentum, $\vec q$, and
the eikonal phase, $\delta$, is given by
\be
\delta(s,\vec b) = \frac{1}{2s}\int\frac{\rd^{2}\vec q}{(2\pi)^{2}}\,e^{-i\vec q\cdot\vec b}{\cal M}_{4}(s, -\vec q\,{}^2)\,,
\label{eq:eikphaseint}
\ee
where ${\cal M}_{4}(s,t)$ is the tree-level $t$-channel amplitude evaluated in the eikonal limit.  When the external  polarizations have spin, there will also be polarization labels on the eikonal phase and amplitudes.

The eikonal phase is related to the time delay experienced by one particle scattering off of the other, in lightcone coordinate time~\cite{Kabat:1992tb,Camanho:2014apa,Hinterbichler:2017qyt}
\be
\Delta x^{\scriptscriptstyle -} = \frac{1}{\lvert p^{\scriptscriptstyle -}\rvert}\delta(s, b).
\ee
In UV-complete theories, the expectation is that particles can only experience time delays $\Delta x^{\scriptscriptstyle -}\geq 0$, never time advances $\Delta x^{\scriptscriptstyle -}< 0$.  This corresponds to a positivity requirement on the eikonal phase, $\delta \geq 0$.

As noted in~\cite{Camanho:2014apa}, the eikonal phase only depends on the on-shell three-point scattering amplitudes of the theory.  This is most easily seen by performing a complex deformation of the $q_1$ integration contour in~\eqref{eq:eikphaseint}.  Upon deformation, the integral picks up the residue of any $t$-channel poles.  By complex factorization, this residue is given by a product of on-shell three point amplitudes~\cite{Schuster:2008nh}.  This results in a contribution
\be
\delta(s,b) = \frac{\sum_I{\cal M}_3^{13 I}(i\partial_{\vec b}){\cal M}_3^{I24}(i\partial_{\vec b})}{2s}\int\frac{\rd^{2}\vec q}{(2\pi)^{2}}\frac{e^{-i\vec q\cdot\vec b}}{\vec q^2+m^2}\, \label{eikonalampde}
\ee
for each exchanged particle of mass $m$.
Here the index $I$ sums over polarizations of the exchanged particle.

If the eikonal phase is negative, $\delta<0$, for some choice of polarizations, it means that there is asymptotic superluminality in the theory visible at a scale $\sim m$, which is the typical scale suppressing derivatives in the cubic vertices in the effective theories of interest.  Assuming the UV theory does not have such asymptotic superluminality, either the cubic vertices responsible for $\delta<0$ must be absent, or new physics must come in at a scale $\sim m$, meaning there cannot be a parametric gap to the next most massive states, as would be required to have an effective field theory with a cutoff parametrically larger than the mass.  Thus, under the assumption of such a UV completion with a gap, we can place sharp model-independent constraints on the cubic vertices.

On-shell cubic vertices are strongly constrained by Lorentz invariance.  In $D=4$ dimensions, there are only 12 possible cubic interaction vertices involving combinations of a single massive spin-2 particle and a single massless spin-2 particle.  Demanding the absence of a time advance in eikonal scattering leaves a particular linear combination of cubic vertices which on-shell are equivalent to the Einstein--Hilbert cubic structure.  Within this linear combination, there is a free parameter which roughly corresponds to the ratio of the Planck masses for the two spin-2 fields.  When both spin-2 particles couple to matter, we find additional constraints on the matter couplings.

One application of these constraints is to ghost-free ``bi-gravity" \cite{Hassan:2011zd}.  This is an effective theory of two interacting spin-2 particles, one massive and one massless, with a cutoff scale parametrically larger than the mass of the massive spin-2 particle.  The theory contains two free parameters in addition to the two Planck masses, the spin-2 mass and a possible cosmological constant for each of the metrics.  Among theories generalizing massive gravity, this theory has seen the widest application to cosmology (see \cite{Schmidt-May:2015vnx,Bull:2015stt,Solomon:2015hja,Hinterbichler:2017sbd} for reviews and further references).   The constraints we find here reduce the two-parameter family of theories to a one-parameter sub-family.  In $D=4$ dimensions, this one-parameter family is the unique choice for which there is a $\mathbb{Z}_2$ symmetry under the interchange of the two metrics. Our constraints also constrain the coupling of ghost-free bi-gravity to matter, as summarized in Figure \ref{fig:constraintplot} for the case of coupling to a single scalar field. We determine these constraints  both through eikonal scattering and by considering classical scattering off of a shockwave background in the ghost-free bimetric theory.

Another application is to large-$N$ confining QCD, which is essentially a weakly-coupled theory of interacting higher spins (the hadrons and glueballs of the theory).   The constraints we derive here also serve as constraints on the possible interactions of a large-$N$ QCD with a massive spin-2 particle as the lightest state, and the way in which it can interact with gravity.  If some large-$N$ QCD-like theory has an isolated massive spin-2 excitation, then its cubic self-couplings and its couplings to gravity must appear in the specific combinations we find.

\section{On-Shell Cubic Amplitudes}
\label{sec:cubic}
The $2\rightarrow 2$ scattering amplitudes in the eikonal regime are only sensitive to the on-shell cubic vertices in a theory.  The first step is therefore to enumerate the possible on-shell cubic vertices for a given particle content.  These three-point vertices are strongly constrained by Lorentz invariance---they are fixed to have a finite number of structures for a given particle content. Here we will use the construction presented in~\cite{Costa:2011mg} to enumerate the possible cubic vertices describing the interactions in the theory of a single massless and a single massive spin-2 particle.

\subsection{General construction}
Consider the on-shell amplitude for the scattering of three particles with integer spins, $s_a$, and masses, $m_a$, where $a = 1,2,3$.  Each particle has a corresponding momentum vector, $p_a^\mu$, which are on-shell, {\it i.e.}, satisfy $p_a^2=-m_a^2$ and $p_1^\mu+p_2^\mu+p_3^\mu=0$.
The polarization data is carried by $\epsilon_a^{\mu_1\ldots \mu_{s_a}}$, which is a symmetric, transverse ($p_{a \mu}\epsilon_a^{\mu\mu_2\ldots \mu_{s_a}} =0$) traceless ($\epsilon_{a \mu}{}^{ \mu\mu_3\ldots \mu_{s_a}} = 0$) tensor associated to each particle.
Since the polarization tensors are symmetric, it is convenient to make the replacement
\be \label{etoz}
\epsilon_a^{\mu_1\cdots \mu_{s_a}}\mapsto z_a^{\mu_1}\cdots z_a^{\mu_{s_a}}, 
\ee
where $z_a^{\mu}$ is some auxiliary vector.\footnote{This is just a trick to keep track of index contractions and does not mean that the physical polarizations can be written as outer products.} The transversality and traceless conditions on $\epsilon^{\mu_1\cdots\mu_{s_a}}$ translate into the condition that $z_a^{\mu}$ is both null and transverse: $z_a^2 = 0$ and $ p_a\cdot z_a = 0$.

Tree amplitudes are Lorentz-invariant contractions of the momenta and polarizations constructed from polynomials of the scalar invariants built by contracting $z$'s and $p$'s. The three types of contraction are $z_{ab} \equiv z_a \cdot z_{b }$, $zp_{ab} \equiv z_a \cdot p_{b}$, and $p_{ab} \equiv p_a \cdot p_{b}$. For cubic amplitudes there are no independent $p_{ab}$ invariants because they can be written in terms of masses using momentum conservation and the on-shell conditions. We can also eliminate three of the six possible $zp_{ab}$ invariants by contracting the momentum conservation equation with each $z_{a}$.
The most general on-shell three-point scattering amplitude can thus be written as a sum of terms of the form
\be \label{generalcubicamp}
{\cal A}_3(p_1,p_2,p_3) \propto  
z_{12}^{n_{12}} z_{13}^{n_{13}} z_{23}^{n_{23}}  zp_{12 }^{m_{12}}  zp_{23 }^{m_{23}}  zp_{31 }^{m_{31}},
\ee
where the exponents $n_{ab}$ and $m_{ab}$ are non-negative integers. 
Since each $z_a$ comes from a polarization tensor \eqref{etoz}, the amplitude \eqref{generalcubicamp} must also be homogeneous of order $s_a$ in each of the $z_a$, meaning that the equations
\begin{subequations} \label{homogeneous}
\begin{align} 
n_{12}+n_{13}+m_{12} &= s_1 ,\\
n_{12}+n_{23}+m_{23} &= s_2 ,\\
n_{13}+n_{23}+m_{31} &= s_3,
\end{align}
\end{subequations}
must be satisfied.
These equations have a finite number of solutions and each solution gives an independent cubic amplitude~\cite{Costa:2011mg,Kravchuk:2016qvl}.
If the process of interest involves identical particles, we can decompose the amplitudes into irreducible representations of the symmetric group which acts by interchanging particles. In what follows, we will restrict our interest to situations where we have a single particle of a given type and so the kinematic factors must be symmetric under exchange of some of the external particles, but in more general situations the kinematic factors can be antisymmetric with the introduction of color factors.

When there are massless particles with spin $\geq 1$, the amplitudes must be gauge invariant.  If particle $a$ is massless, this corresponds to the invariance of the amplitude under the replacement
\be
z_a \mapsto z_a + \varepsilon p_a\, ,
\label{eq:gaugeinv}
\ee
which generally reduces the number of allowed structures. (There can also be parity odd cubic amplitudes in five or fewer dimensions, which contain contractions of $z$'s and $p$'s with an antisymmetric tensor, but here we will restrict our attention to the parity-even cases.)

In four or fewer dimensions there can also be redundancies in the cubic amplitudes due to dimensionally dependent identities. For example, in four-dimensions, which is the case we will be mostly interested in, any five four-vectors cannot be linearly independent.  This implies that the following Gram determinant of the five vectors $z_1$, $z_2$, $z_3$, $p_1$ and $p_2$ must vanish~\cite{Costa:2011mg,Conde:2016izb}:
\be
\label{gramidentity}
\epsilon^{\mu_1\mu_2\mu_3\mu_4\mu_5}\epsilon_{\nu_1\nu_2\nu_3\nu_4\nu_5} p_{1\,\mu_1}p_1^{\nu_1}p_{2\,\mu_2}p_2^{\nu_2}z_{1\,\mu_3}z_1^{\nu_3}z_{2\,\mu_4}z_2^{\nu_4}z_{3\,\mu_5}z_3^{\nu_5} = 0.
\ee
This gives relations between the various structures built from these constituents.

We now proceed to list the three-point amplitudes that will be relevant for the eikonal scattering calculation. These are obtained by first solving \eqref{homogeneous} and then imposing the relevant particle exchange/gauge symmetries on the resulting amplitudes \eqref{generalcubicamp}. In what follows there are two mass scales: $m$ is the mass of the massive spin-2 particle, which is used to suppress powers of the derivatives in the interaction vertices, and $M_{\rm Pl}$ is the Planck mass, which will suppress powers of the field.
\subsection{Three massive particles}
Solving the equations \eqref{homogeneous} for the case of three identical massive spin-2 particles gives five independent cubic vertices which are symmetric under interchange:

\begin{itemize}
\item Zero derivatives
\be
\label{eq:3massivespin20d}
\mathcal{A}_1 =  m^2 z_{12} z_{13} z_{23},
\ee
\item Two derivatives
\begin{subequations}
\begin{align}
\mathcal{A}_2 &= z_{23}^2 zp_{12}^2+z_{13}^2 zp_{23}^2+z_{12}^2 zp_{31}^2, \\
\mathcal{A}_3 &= z_{13}z_{23} zp_{12}zp_{23}+z_{12}z_{23}zp_{12}zp_{31}+z_{12}z_{13}zp_{23}zp_{31}, 
\end{align}
\end{subequations}
\item Four derivatives 
\be
\mathcal{A}_4 = \frac{1}{m^2} zp_{12} zp_{23} zp_{31} \left( z_{12}zp_{31}+z_{23} zp_{12} +z_{13} zp_{23} \right),
\ee
\item Six derivatives
\be
\label{eq:3massivespin26d}
\mathcal{A}_5  = \frac{1}{m^4}  zp_{12}^2 zp_{23}^2 zp_{31}^2.
\ee
\end{itemize}
The most general interaction vertex is then given by a linear combination of these structures as\footnote{To obtain the $D$-dimensional vertices send $M_{\rm Pl} \mapsto M_{\rm Pl}^{(D-2)/2}$.}
\be
\mathcal{V}_a = \frac{i}{M_{\rm Pl}} \sum_{i=1}^5 a_i \mathcal{A}_i,
\ee
where $a_i$ are dimensionless constants.

It is worth noting that the structures~\eqref{eq:3massivespin20d}--\eqref{eq:3massivespin26d} do not directly correspond to the on-shell amplitudes 
generated by a fiducial basis of Lagrangian structures. For example, the amplitude from the Einstein--Hilbert cubic vertex is a linear combination of ${\cal A}_1, {\cal A}_2$ and ${\cal A}_3$, and another linear combination is the amplitude from the 2-derivative non-gauge invariant ``pseudo-linear'' term of \cite{Folkerts:2011ev,Hinterbichler:2013eza}. This is relevant because the four-dimensional identity~\eqref{gramidentity} in this case reads
\be
4 \mathcal{A}_4 -2\mathcal{A}_2-2\mathcal{A}_3 +3 \mathcal{A}_1 =0,
\ee
which corresponds precisely to the vanishing of the Gauss--Bonnet term in four dimensions.\footnote{Note that even though the Gauss-Bonnet term has no non-derivative terms, a part proportional to ${\cal A}_1$ will be generated by the massive on-shell kinematics.} We therefore see that there are actually only four linearly independent structures in $D=4$. For a precise translation between the structure basis we adopt here and a fiducial Lagrangian basis, see~\cite{Hinterbichler:2017qyt}.
\subsection{Two massive particles and one massless particle}
\label{eq:2massive1massless}
Next we consider three-point amplitudes where two identical massive spin-2 particles and a massless spin-2 particle interact.  Taking particle 3 to be the massless particle, there are six possible structures symmetric under interchanging particles $1$ and $2$,
\begin{itemize}
\item Two derivatives
\begin{subequations}
\begin{align}
\mathcal{B}_1 &= z_{12}^2 zp_{31}^2, \\
\mathcal{B}_2 &= z_{12} zp_{31}\left(z_{23} zp_{12} + z_{1 3} zp_{2 3}\right),  \\
\mathcal{B}_3 &= (z_{2 3} zp_{1 2}+ z_{1 3} zp_{2 3})^2, 
\end{align}
\end{subequations}

\item Four derivatives
\begin{subequations}
\begin{align}
\mathcal{B}_4 &= \frac{1}{m^2} zp_{12} zp_{2 3} zp_{3 1}\left(z_{2 3} zp_{1 2} + z_{13} zp_{2 3}\right) ,\\
\mathcal{B}_5 &=  \frac{1}{m^2} z_{1 2} zp_{1 2} zp_{2 3} zp_{31}^2,
\end{align}
\end{subequations}

\item Six derivatives
\be
\mathcal{B}_6 =  \frac{1}{m^4} zp_{12}^2 zp_{23}^2 zp_{31}^2.
\ee
\end{itemize}
A general interaction vertex is a linear combination of these structures
\be
\mathcal{V}_b = \frac{i}{M_{\rm Pl}} \sum_{i=1}^6 b_i \mathcal{B}_i,
\ee
where $b_i$ are dimensionless constants.
In this case, the four-dimensional identity \eqref{gramidentity} becomes
\be
2{\cal B}_5+2 \mathcal{B}_4 -\mathcal{B}_3 =0\, ,
\ee
so we see that there are only five linearly-independent such structures in $D=4$.

\subsection{Two massless particles and one massive particle}
Next we consider the interactions between two identical massless particles and one massive particle. Taking particle 3 to be the massive particle, there are two possible structures symmetric under interchanging particles $1$ and $2$,%
\begin{itemize}
\item Four derivatives
\be
\mathcal{C}_1 = \frac{1}{m^2}  \left(m^2 z_{1 2} - 2 zp_{1 2} zp_{2 3}\right) \left(m^2 z_{1 3} z_{2 3} + 
   2 zp_{3 1} \left(z_{2 3} zp_{1 2} + z_{1 3} zp_{2 3} + 
      z_{1 2} zp_{3 1}\right) \right),  
\ee

\item Six derivatives
\be
\mathcal{C}_2 =  \frac{1}{m^4} \left(m^2 z_{1 2} - 2 zp_{1 2} zp_{2 3}\right)^2 zp_{3 1}^2.
\ee
\end{itemize}
A general interaction vertex is then
\be
\mathcal{V}_c = \frac{i}{M_{\rm Pl}} \sum_{i=1}^2 c_i \mathcal{C}_i,
\ee
where $c_i$ are dimensionless constants.
The four-dimensional identity \eqref{gramidentity} in this case becomes
\be
\mathcal{C}_1=0,
\ee
so that in $D=4$ only the ${\cal C}_2$ structure survives.

\subsection{Three massless particles}

Finally we consider the case when all three particles are identical and massless. There are three possible structures symmetric under interchange:
\begin{itemize}
\item Two derivatives
\be
\label{eq:3massless1}
\mathcal{D}_1 = \left(z_{2 3} zp_{1 2} + z_{1 3} zp_{2 3} + z_{12} zp_{3 1}\right)^2,
\ee
\item Four derivatives
\be
\mathcal{D}_2 = \frac{1}{m^2} zp_{1 2} zp_{2 3} zp_{31} \left(z_{2 3} zp_{1 2} + z_{13} zp_{23} + z_{1 2} zp_{3 1}\right),
\ee
\item Six derivatives
\be
\mathcal{D}_3 = \frac{1}{m^4} zp_{12}^2 zp_{23}^2 zp_{31}^2. 
\label{eq:3massless3}
\ee
\end{itemize}
The general interaction vertex is constructed as
\be
\mathcal{V}_d = \frac{i}{M_{\rm Pl}} \sum_{i=1}^3 d_i \mathcal{D}_i,
\ee
where $d_i$ are dimensionless constants.
The $D=4$ identity~\eqref{gramidentity} now implies
\be
\mathcal{D}_2=0,
\ee
which corresponds to the vanishing of the Gauss--Bonnet term in four dimensions.

In the case of massless particles, there is a straightforward relation between the structures~\eqref{eq:3massless1}--\eqref{eq:3massless3} and the amplitudes which arise from simple Lagrangian terms. Specifically, ${\cal D}_1$ is proportional to the Einstein--Hilbert amplitude, ${\cal D}_2$ is proportional to the Gauss--Bonnet cubic term's amplitude and ${\cal D}_3$ is proportional to the cubic amplitude arising from $R_{\mu\nu\rho\sigma}^3$.

\subsection{Scalar coupling}
We will also be interested in the eikonal amplitude where massive and massless spin-2 particles interact with a scalar source, so we will require the 
 cubic amplitudes involving one spin-2 particle and two scalar particles. There is a unique such vertex coupling a spin-2 particle with two identical scalars, irrespective of whether the spin-2 is massive or massless:
\be
{\cal K} = zp_{31}^2,
\ee
where particle 3 has spin 2. The vertex coupling the scalar to the massless spin-2 particle is then
\be
\mathcal{V}^{m=0}_{\kappa} = \frac{i}{M_{\rm Pl}} \kappa_0 {\cal K},
\ee
and the vertex coupling the scalar to the massive spin-2 particle is
\be
\mathcal{V}^{m\neq0}_{\kappa} = \frac{i}{M_{\rm Pl}} \kappa_m {\cal K},
\ee
where $\kappa_0, \kappa_m$ are constants.

We could also consider cubic amplitudes involving one scalar and two spin-2 particles but these would require the exchange of a scalar in the eikonal amplitude, which is suppressed relative to spin-2 exchange at large center-of-mass energy.

\subsection{$S$-matrix equivalence principle constraints}
\label{sec:smatrixequivalence}
So far we have allowed the possible cubic amplitudes to have arbitrary coefficients. However, for cubic vertices involving 
massless spin-2 fields there are constraints coming from the decoupling of longitudinal polarizations in higher-point amplitudes~\cite{Weinberg:1965nx}. 
These constraints follow from Poincar\'e invariance and the masslessness of the graviton and are the $S$-matrix manifestation of the equivalence principle. The requirement is that any particle that couples directly or indirectly to Einstein gravity must have a gravitational minimal coupling interaction with a universal coupling constant~\cite{Porrati:2012rd}.

The $S$-matrix equivalence principle imposes the following constraints on the cubic couplings  depending on what couplings to the massless spin-2 particle are nonzero:
\begin{enumerate}[label=\textbf{\arabic*}.]
\item When $b_1 \neq 0$ and $\kappa_0 \neq 0$ the equivalence principle implies $b_1=\kappa_0=d_1$,

\vspace{.25cm}
\item When $b_1 \neq 0$ and $\kappa_0 = 0$ then we have $\kappa_m=0$ and $b_1 =d_1$,

\vspace{.25cm}
\item When $b_1 =0$ and $\kappa_0 \neq 0$ then we have $b_i=\kappa_m=0$ and $\kappa_0=d_1$,

\vspace{.25cm}
\item When $b_1 =0$ and $\kappa_0 =0$ then $b_i=0$.
\end{enumerate}
The second case ($b_1\neq 0, \kappa_0 = 0$) is when the scalar decouples from gravity, the third case ($b_1 = 0, \kappa_0\neq 0$) is when the massive spin-2 particle decouples from gravity, and the last case ($b_1 = \kappa_0= 0$)  is when both decouple. In GR the usual canonically-normalized gravitational coupling corresponds to $d_1=2$, which can always be set by a rescaling of $M_{\rm Pl}$ if $d_1 \neq 0$.

We emphasize that the $S$-matrix equivalence principle constraints, although powerful, require only very weak assumptions, namely the masslessness of the graviton and Poincar\'e invariance. When we impose these constraints we are thus not making any assumptions beyond those already implicitly made. 

\section{Eikonal Scattering in $D=4$}
\label{sec:scalarspin2}
The eikonal regime consists of two highly boosted particles moving approximately along orthogonal null directions in flat space. The kinematics in this situation are somewhat nonstandard, so here we describe a convenient parameterization of the momenta and a basis for the polarizations of the external particles.

\subsection{Kinematics}

The interactions listed in Section~\ref{sec:cubic} allow for mixing between massive and massless spin-2 particles on the rails of the ladder diagrams for eikonal scattering (see \eqref{eq:feyndiagrams} below), so we need to generalize the eikonal kinematics used in~\cite{Hinterbichler:2017qyt} to allow for this. We work in lightcone coordinates $(x^{\scriptscriptstyle -},x^{\scriptscriptstyle +},x^i)$ defined by
\be
\label{eq:lconecoords}
 x^{\scriptscriptstyle \pm}={1\over \sqrt{2}}\left(x^0\pm x^1\right)\, .
\ee
In these coordinates, the Minkowski line element takes the form
\be
\rd s^2 = -2\rd x^{\scriptscriptstyle +}\rd x^{\scriptscriptstyle -}+\delta_{ij}x^i x^j,
\ee
where $\delta_{ij}$ is the $2\times2$ identity matrix.

The external momenta are given by
\begin{subequations}
\begin{align}
p_1^\mu&=\left({1\over 2p^{\scriptscriptstyle +}}\left({{\vec q\ }^2\over 4}+m_1^2\right),p^{\scriptscriptstyle +} ,{ q^i \over 2}\right)\, , & p_3^\mu&=\left({1\over 2 \tilde{p}^{\scriptscriptstyle +}}\left({{\vec q\ }^2\over 4}+m_3^2\right),\tilde{p}^{\scriptscriptstyle +},-{ q^i \over 2}\right)\, , \\
p_2^\mu&=\left(p^{\scriptscriptstyle -},{1\over 2 p^{\scriptscriptstyle -}}\left({{\vec q\ }^2\over 4}+m_2^2\right), -{ q^i\over 2 }\right)\, , & p_4^\mu &=\left(\tilde{p}^{\scriptscriptstyle -},{1\over 2 \tilde{p}^{\scriptscriptstyle -}}\left( {{\vec q\ }^2\over 4}+m_4^2\right),{ q^i\over 2 }\right)\,.
\end{align}
\end{subequations}
Here $p^{\scriptscriptstyle +}$ and $p^{\scriptscriptstyle -}$ are the independent lightcone momenta which are taken to be large in the eikonal limit, and the constants $\tilde{p}^{\scriptscriptstyle +}$ and $\tilde{p}^{\scriptscriptstyle -}$ are fixed by imposing momentum conservation. There are always two solutions for $\tilde{p}^{\scriptscriptstyle +}$ and $\tilde{p}^{\scriptscriptstyle -}$ consistent with momentum conservation. If all the particle masses are equal, then one solution is 
\be \label{equalmasssol}
\tilde{p}^{\scriptscriptstyle +} = p^{\scriptscriptstyle +}, \quad \tilde{p}^{\scriptscriptstyle -} = p^{\scriptscriptstyle -},
\ee
which gives the usual eikonal kinematics---as in~\cite{Hinterbichler:2017qyt}---and the second solution gives $p^{\mu}_1=p^{\mu}_4$ and  $p^{\mu}_2=p^{\mu}_3$, which corresponds to backward scattering. When the masses are different we choose the solution that at high energies reduces to \eqref{equalmasssol} in the limit of equal masses.\footnote{We match solutions at large energy because the equal-mass limit of one of the general solutions gives~\eqref{equalmasssol} at high energies and backward scattering at low energies, while the other general solution gives the opposite. This is possible because the equal mass solutions coincide when $s=4m^2+q^2$ and so the space of solutions is not smooth. We will not write out the explicit expressions for $\tilde p^{\scriptscriptstyle +}$ and $\tilde p^{\scriptscriptstyle -}$, as they are rather long and not particularly enlightening. The solution that in the equal-mass limit reduces to \eqref{equalmasssol} at high energies is the correct one for eikonal scattering since we always work in the high-energy limit.} The Mandelstam variables are given by
\begin{align}
s&= -(p_1+p_2)^2=\frac{\left(8  p^{\scriptscriptstyle +} p^{\scriptscriptstyle -} + {\vec q\ }^2+ 4 m_1^2\right) \left(8  p^{\scriptscriptstyle +} p^{\scriptscriptstyle -} + {\vec q\ }^2 + 4 m_2^2\right)}{32 p^{\scriptscriptstyle +}  p^{\scriptscriptstyle -}}\simeq 2 p^{\scriptscriptstyle +}p^{\scriptscriptstyle -}\, ,\\
t&= -(p_1-p_3)^2= -\frac{\left(p^{\scriptscriptstyle +} +\tilde{p}^{\scriptscriptstyle +}\right)^2 {\vec q\ }^2 + 
  4 \left(p^{\scriptscriptstyle +} - \tilde{p}^{\scriptscriptstyle +}\right) \left( p^{\scriptscriptstyle +} m_3^2-\tilde{p}^{\scriptscriptstyle +} m_1^2 \right)}{4 p^{\scriptscriptstyle +} \tilde{p}^{\scriptscriptstyle +}}\simeq -\vec q\,{}^2\, .
\end{align}
The last expressions are their approximate form in the high energy eikonal limit where $p^{\scriptscriptstyle +},p^{\scriptscriptstyle -}\rightarrow \infty$.

\subsection{Polarizations}
We also need a basis for the polarization vectors of the external particles. These are given by
\begin{subequations}\label{spi1polse}
\begin{align}
 \epsilon_{T, \lambda}^\mu(p_1)&=\left({{\vec q \ }\cdot {\vec e}_{ \lambda} \over 2p^{\scriptscriptstyle +}},0 ,{\vec e}_{\lambda}\right)\, , &  \epsilon_L^\mu(p_1)&=\left({1\over 2m_1p^{\scriptscriptstyle +}}\left({{{\vec q \ }}^2\over 4}-m_1^2\right),{p^{\scriptscriptstyle +}\over m_1} ,{ q^i \over 2m_1}\right)\, ,\\
 \epsilon_{T, \lambda}^\mu(p_2)&=\left(0,-{{\vec q \ }\cdot  {\vec e}_{ \lambda} \over 2p^{\scriptscriptstyle -}},{\vec e}_{ \lambda}\right)\, ,&  \epsilon_L^\mu(p_2)&=\left({p^{\scriptscriptstyle -}\over m_2} ,{1\over 2m_2p^{\scriptscriptstyle -}}\left({{{\vec q \ }}^2\over 4}-m_2^2\right),-{ q^i \over 2m_2}\right)\, , \\
\epsilon_{T, \lambda}^\mu(p_3)&=\left(-{{\vec q \ }\cdot {\vec e}_{ \lambda} \over 2\tilde{p}^{\scriptscriptstyle +}},0 ,{\vec e}_{ \lambda}\right)\, ,  &\epsilon_L^\mu(p_3)&=\left({1\over 2m_3 \tilde{p}^{\scriptsize +}}\left({{{\vec q \ }}^2\over 4}-m_3^2\right), \frac{\tilde{p}^{\scriptsize +}}{m_3} ,-{ q^i \over 2m_3}\right)\, ,  \\
 \epsilon_{T, \lambda}^\mu(p_4)&=\left(0,{{\vec q \ }\cdot {\vec e}_{ \lambda} \over 2\tilde{p}^{\scriptscriptstyle -}} ,{\vec e}_{ \lambda}\right)\, , &  \epsilon_L^\mu(p_4)&=\left(\frac{\tilde{p}^{\scriptsize -}}{m_4},{1\over 2m_4\tilde{p}^{\scriptsize -}}\left({{{\vec q \ }}^2\over 4}-m_4^2\right) ,{ q^i \over 2m_4}\right)\, ,
\end{align}
\end{subequations}
where the vectors ${\vec e}_{ \lambda}$ form an orthonormal basis for the $2$-dimensional transverse plane,
\be
e_{ \lambda i}  e^j_{ \lambda'} = \delta_{\lambda \lambda'}, \quad \sum_{\lambda} e^i_{ \lambda} e^j_{ \lambda} = \delta^{ij}.
\ee
A convenient basis, which we will use, is to take $e^i_{\lambda} = \delta^i_{\lambda}$.
The polarization vectors \eqref{spi1polse} are all transverse to their momenta, orthonormal, and they satisfy the completeness relation
\be
\epsilon_{L}^\mu (p_a) \epsilon_L^\nu(p_a)^\ast+\sum_\lambda \epsilon_{T,\lambda}^\mu (p_a) \epsilon_{T,\lambda}^\nu(p_a)^\ast=\eta^{\mu\nu}-{1\over p_a^2}p_a^\mu p_a^\nu\, ,
\ee
where $a=1,2,3,4$ labels the four particles.

We can use the polarization vectors~\eqref{spi1polse} to construct polarization tensors for the external spin-2 fields.  For a massive spin-2 particle we choose the polarizations given by 
\begin{subequations} \label{spin2polse}
\begin{align}
\epsilon_{T, \tilde{\lambda}}^{\mu\nu}(p_a) &=\epsilon_{T, \lambda}^{\mu}(p_a)\epsilon_{T, \lambda'}^{\nu}(p_a)\, , \\
\epsilon_{V, \lambda} ^{\mu\nu}(p_a) &={i\over \sqrt{2}}\left(\epsilon_{T, \lambda}^{\mu}(p_a)\epsilon_L^{\nu}(p_a) +\epsilon_L^{\mu}(p_a)\epsilon_{T, \lambda}^{\nu}(p_a) \right)\, ,\\
 \epsilon_S^{\mu\nu}(p_a) &= \sqrt{3\over 2}\left[ \epsilon_L^{\mu}(p_a) \epsilon_L^{\nu}(p_a) -{1\over 3}\left(\eta^{\mu\nu} -{1\over p_a^2}p_a^\mu p_a^\nu  \right) \right] \, , 
\end{align}
\end{subequations}
where $T$, $V$, and $S$ stand for tensor, vector and scalar polarizations, respectively.
In the expression for $\epsilon_{T, \tilde{\lambda}}^{\mu\nu}(p_a) $ we replace $ e^i_{\lambda} e^j_{\lambda'} \mapsto e^{ij}_{\tilde{\lambda}}$, where the $e^{ij}_{\tilde{\lambda}}$ are two-tensors that form an orthonormal basis of symmetric traceless tensors,
\be
e_{ij, \tilde{\lambda}} e^{ij}_{\tilde{\lambda}'}=\delta_{\tilde{\lambda} \tilde{\lambda}'}, \quad \sum_{\tilde{\lambda}} e^{ij}_{\tilde{\lambda}} e^{kl}_{\tilde{\lambda}} =\frac{1}{2} \left( \delta^{ik} \delta^{jl} + \delta^{jk} \delta^{il} - \delta^{ij} \delta^{kl} \right).
\ee
Explicitly, we take the two tensors to be the standard plus and cross polarizations,
\be
{\bf e}_{\oplus}=  \frac{1}{\sqrt{2}}\left(\begin{array}{cc}1 & 0  \\0 & -1 \end{array}\right), \quad {\bf e}_{\otimes}=  \frac{1}{\sqrt{2}}\left(\begin{array}{cc}0 & 1  \\1 & 0 \end{array}\right),
\ee
which means that the full tensor polarizations are
\begin{align}
\epsilon_{T, \oplus}^{\mu\nu}&=\frac{1}{\sqrt{2}}\left( \epsilon_{T, 1}^{\mu}\epsilon_{T, 1}^{\nu}-\epsilon_{T, 2}^{\mu}\epsilon_{T, 2}^{\nu} \right) , \\
\epsilon_{T, \otimes}^{\mu\nu}&=\frac{1}{\sqrt{2}}\left( \epsilon_{T, 1}^{\mu}\epsilon_{T, 2}^{\nu}+\epsilon_{T, 2}^{\mu}\epsilon_{T, 1}^{\nu} \right).
\end{align}
For a massless spin-2 particle we have only these tensor polarizations.

The polarization tensors so defined are all transverse and orthonormal. Additionally they satisfy the completeness relations
\be
\label{eq:massiveprop}
 \epsilon_{S}^{\mu\nu} (p_a) \epsilon_S^{\alpha\beta}(p_a)^\ast+\sum_\lambda \epsilon_{V,\lambda}^{\mu\nu} (p_a) \epsilon_{V,\lambda}^{\alpha\beta}(p_a)^\ast+\sum_{\tilde\lambda}\epsilon_{T,\tilde \lambda}^{\mu\nu} (p_a) \epsilon_{T,\tilde\lambda}^{\alpha\beta}(p_a)^\ast ={1\over 2}\left( P^{\mu\alpha} P^{\nu\beta}+P^{\nu\alpha} P^{\mu\beta}\right)-{1\over 3}P^{\mu\nu} P^{\alpha\beta},
\ee
with $P_{\mu\nu} = \eta_{\mu\nu}+\frac{1}{m^2}p_\mu p_\nu$, and
\be
\label{eq:masslessprop}
\sum_{\tilde\lambda}\epsilon_{T,\tilde \lambda}^{\mu\nu} (p_a) \epsilon_{T,\tilde\lambda}^{\alpha\beta}(p_a)^\ast ={1\over 2}\left( \eta^{\mu\alpha} \eta^{\nu\beta}+\eta^{\nu\alpha} \eta^{\mu\beta}-\eta^{\mu\nu} \eta^{\alpha\beta}\right).
\ee
The right hand sides of equations~\eqref{eq:massiveprop} and~\eqref{eq:masslessprop} are the massive spin-2 and massless spin-2 (in de Donder gauge) propagator numerators, respectively.

\subsection{Scalar--Spin-2 eikonal scattering}
\label{eq:d4scalarspin2}
After these preliminaries, we are now ready to describe the computation of the eikonal scattering amplitude. The object that we are interested in is the eikonal phase 
\be
\label{eq:eikphase}
\delta(s,\vec b) = \frac{1}{2s}\int\frac{\rd^{2}\vec q}{(2\pi)^{2}}\,e^{-i\vec q\cdot\vec b}{\cal M}_{4}(s, -\vec q\,{}^2),
\ee
where ${\cal M}_4(s,t)$ is the tree-level $t$-channel amplitude in the eikonal limit $p^{\scriptscriptstyle +},p^{\scriptscriptstyle -}\rightarrow \infty$.  We begin by considering eikonal scattering between a scalar source particle and an admixture of massless and massive spin-2 particles, which proceeds through the exchange of both massive and massless spin-2 modes.

We take the vertices to be the general linear combinations $\mathcal{V}_{a}$, $\mathcal{V}_{b}$, $\mathcal{V}_{c}$, $\mathcal{V}_{d}$, $\mathcal{V}^{m=0}_{\kappa}$, and $\mathcal{V}^{m \neq 0}_{\kappa}$ as defined in Section \ref{sec:cubic}. In this section we concern ourselves with the eikonal amplitude in $D=4$, which allows us to use the identity~\eqref{gramidentity}  to set $a_4=b_4=c_1=d_2=0$ without loss of generality. (We will comment on the case $D>4$ in Section~\ref{sec:dneq4amplitudes}.) Altogether there are are 8 inequivalent diagrams that contribute,%
\vspace{-.5cm}
\begin{align}
\nonumber
&\begin{tikzpicture}[line width=1.7 pt,baseline={([yshift=-3ex]current bounding box.center)},vertex/.style={anchor=base,
    circle,fill=black!25,minimum size=18pt,inner sep=2pt}]
\draw[style] (.5,0) -- (2.5,0);
\draw[style={decorate,decoration=complete sines},line width=1] (0.5,1.05) -- (4.5,1.05);
\draw[style={decorate,decoration=complete sines},line width=1] (1.5,0) -- (1.5,2);
\node[scale=1] at (.2, -.02) {$p_2$};
\node[scale=1] at (2.8, -.02) {$p_4$};
\node[scale=1] at (.2, 1.06) {$p_1$};
\node[scale=1] at (2.8, 1.06) {$p_3$};
\node[scale=1] at (3.75,.5) {$+$};
\end{tikzpicture}
&&\!\!\!\!\!\!\!\!
\begin{tikzpicture}[line width=1.7 pt,baseline={([yshift=-3ex]current bounding box.center)},vertex/.style={anchor=base,
    circle,fill=black!25,minimum size=18pt,inner sep=2pt}]
\draw[style] (.5,0) -- (2.5,0);
\draw[style={decorate,decoration=complete sines},line width=1] (0.5,1.05) -- (4.5,1.05);
\draw[style={decorate,decoration=complete sines},line width=1] (1.46,0) -- (1.46,2);
\draw[style={decorate,decoration=complete sines},line width=1] (1.54,0) -- (1.54,2);
\node[scale=1] at (.2, -.02) {$p_2$};
\node[scale=1] at (2.8, -.02) {$p_4$};
\node[scale=1] at (.2, 1.06) {$p_1$};
\node[scale=1] at (2.8, 1.06) {$p_3$};
\node[scale=1] at (3.75,.5) {$+$};
\end{tikzpicture}
&&\!\!\!\!\!\!\!\!
\begin{tikzpicture}[line width=1.7 pt,baseline={([yshift=-3ex]current bounding box.center)},vertex/.style={anchor=base,
    circle,fill=black!25,minimum size=18pt,inner sep=2pt}]
\draw[style] (.5,0) -- (2.5,0);
\draw[style={decorate,decoration=complete sines},line width=1] (0.5,1.05) -- (4.5,1.05);
\draw[style={decorate,decoration=complete sines},line width=1] (0.5,1.13) -- (4.5,1.13);
\draw[style={decorate,decoration=complete sines},line width=1] (1.46,0) -- (1.46,2);
\draw[style={decorate,decoration=complete sines},line width=1] (1.54,0) -- (1.54,2);
\node[scale=1] at (.2, -.02) {$p_2$};
\node[scale=1] at (2.8, -.02) {$p_4$};
\node[scale=1] at (.2, 1.06) {$p_1$};
\node[scale=1] at (2.8, 1.06) {$p_3$};
\node[scale=1] at (3.75,.5) {$+$};
\end{tikzpicture}
&&\!\!\!\!\!\!\!\!
\begin{tikzpicture}[line width=1.7 pt,baseline={([yshift=-3ex]current bounding box.center)},vertex/.style={anchor=base,
    circle,fill=black!25,minimum size=18pt,inner sep=2pt}]
\draw[style] (.5,0) -- (2.5,0);
\draw[style={decorate,decoration=complete sines},line width=1] (0.5,1.05) -- (4.5,1.05);
\draw[style={decorate,decoration=complete sines},line width=1] (0.5,1.13) -- (4.5,1.13);
\draw[style={decorate,decoration=complete sines},line width=1] (1.5,0) -- (1.5,2);
\node[scale=1] at (.2, -.02) {$p_2$};
\node[scale=1] at (2.8, -.02) {$p_4$};
\node[scale=1] at (.2, 1.06) {$p_1$};
\node[scale=1] at (2.8, 1.06) {$p_3$};
\end{tikzpicture}\\
&
\begin{tikzpicture}[line width=1.7 pt,baseline={([yshift=-3ex]current bounding box.center)},vertex/.style={anchor=base,
    circle,fill=black!25,minimum size=18pt,inner sep=2pt}]
\draw[style] (.5,0) -- (2.5,0);
\draw[style={decorate,decoration=complete sines},line width=1] (0.5,1.05) -- (2.5,1.05);
\draw[style={decorate,decoration=complete sines},line width=1] (1.5,1.05) -- (3.5,1.05);
\draw[style={decorate,decoration=complete sines},line width=1] (1.5,1.13) -- (3.5,1.13);
\draw[style={decorate,decoration=complete sines},line width=1] (1.5,0) -- (1.5,2);
\node[scale=1] at (.2, -.02) {$p_2$};
\node[scale=1] at (2.8, -.02) {$p_4$};
\node[scale=1] at (.2, 1.06) {$p_1$};
\node[scale=1] at (2.8, 1.06) {$p_3$};
\node[scale=1] at (3.75,.5) {$+$};
\end{tikzpicture}
&&\!\!\!\!\!\!\!\!
\begin{tikzpicture}[line width=1.7 pt,baseline={([yshift=-3ex]current bounding box.center)},vertex/.style={anchor=base,
    circle,fill=black!25,minimum size=18pt,inner sep=2pt}]
\draw[style] (.5,0) -- (2.5,0);
\draw[style={decorate,decoration=complete sines},line width=1] (0.5,1.05) -- (2.5,1.05);
\draw[style={decorate,decoration=complete sines},line width=1] (0.5,1.13) -- (2.5,1.13);
\draw[style={decorate,decoration=complete sines},line width=1] (1.5,1.05) -- (3.5,1.05);
\draw[style={decorate,decoration=complete sines},line width=1] (1.5,0) -- (1.5,2);
\node[scale=1] at (.2, -.02) {$p_2$};
\node[scale=1] at (2.8, -.02) {$p_4$};
\node[scale=1] at (.2, 1.06) {$p_1$};
\node[scale=1] at (2.8, 1.06) {$p_3$};
\node[scale=1] at (3.75,.5) {$+$};
\end{tikzpicture}
&&\!\!\!\!\!\!\!\!
\begin{tikzpicture}[line width=1.7 pt,baseline={([yshift=-3ex]current bounding box.center)},vertex/.style={anchor=base,
    circle,fill=black!25,minimum size=18pt,inner sep=2pt}]
\draw[style] (.5,0) -- (2.5,0);
\draw[style={decorate,decoration=complete sines},line width=1] (0.5,1.05) -- (2.5,1.05);
\draw[style={decorate,decoration=complete sines},line width=1] (1.5,1.05) -- (3.5,1.05);
\draw[style={decorate,decoration=complete sines},line width=1] (1.5,1.13) -- (3.5,1.13);
\draw[style={decorate,decoration=complete sines},line width=1] (1.46,0) -- (1.46,2);
\draw[style={decorate,decoration=complete sines},line width=1] (1.54,0) -- (1.54,2);
\node[scale=1] at (.2, -.02) {$p_2$};
\node[scale=1] at (2.8, -.02) {$p_4$};
\node[scale=1] at (.2, 1.06) {$p_1$};
\node[scale=1] at (2.8, 1.06) {$p_3$};
\node[scale=1] at (3.75,.5) {$+$};
\end{tikzpicture}
&&\!\!\!\!\!\!\!\!
\begin{tikzpicture}[line width=1.7 pt,baseline={([yshift=-3ex]current bounding box.center)},vertex/.style={anchor=base,
    circle,fill=black!25,minimum size=18pt,inner sep=2pt}]
\draw[style] (.5,0) -- (2.5,0);
\draw[style={decorate,decoration=complete sines},line width=1] (0.5,1.05) -- (2.5,1.05);
\draw[style={decorate,decoration=complete sines},line width=1] (0.5,1.13) -- (2.5,1.13);
\draw[style={decorate,decoration=complete sines},line width=1] (1.5,1.05) -- (3.5,1.05);
\draw[style={decorate,decoration=complete sines},line width=1] (1.46,0) -- (1.46,2);
\draw[style={decorate,decoration=complete sines},line width=1] (1.54,0) -- (1.54,2);
\node[scale=1] at (.2, -.02) {$p_2$};
\node[scale=1] at (2.8, -.02) {$p_4$};
\node[scale=1] at (.2, 1.06) {$p_1$};
\node[scale=1] at (2.8, 1.06) {$p_3$};
\end{tikzpicture}
\label{eq:feyndiagrams}
\end{align}
A single wavy line represents a massless spin-2 particle, a double wavy line a massive spin-2 particle, and a straight solid line a scalar.

For the external polarization states for the massive particles we take a linear combination of the polarizations~\eqref{spin2polse},
\be
\epsilon^{(m)}_{\mu\nu}{}^a = P_{a,S} \epsilon^{a,S}_{\mu\nu}+P_{a,V} \epsilon^{a,V}_{\mu\nu}+P_{a,T} \epsilon^{a,T}_{\mu\nu},
\label{eq:massivegeneralpol}
\ee
while for the massless spin-2 particle there is only tensor polarizations,
\be
\epsilon^{(0)}_{\mu\nu}{}^a =Q_{a,T} \epsilon^{a,T}_{\mu\nu}.
\label{eq:maslesspol}
\ee
Here the $P_a, Q_a$ are just coefficients labeling the mixture of different polarization states. It is convenient to assemble all of these polarization coefficients into a big unit-norm vector
\be
\label{eq:Pa}
{\bf P}_a=\left(\begin{array}{c}
P_{a,S}  \\
 P_{a,V} \\
 P_{a,T} \\
 Q_{a,T}
\end{array}
\right) \,,
\ee
so that a general incoming state can be thought of as an admixture of all possible massive and massless polarization states (and the same for an outgoing state). The amplitude is then a transition matrix between these vectors of polarization coefficients.

A somewhat tedious computation of the $t$ channel diagrams~\eqref{eq:feyndiagrams} in the eikonal limit yields the following form for the eikonal phase after transforming to impact parameter space:\footnote{The Fourier transform with respect to the momentum transfer, $\vec q$ requires the following formulae:
\be
\int\frac{\rd^{2}\vec q}{(2\pi)^{2}}\frac{e^{-i\vec q\cdot\vec b}}{\vec q^2 + m^2} = \frac{1}{2\pi}K_0(mb),~~~~~ {\rm and}~~~~~\int\frac{\rd^{2}\vec q}{(2\pi)^{2}}\frac{e^{-i\vec q\cdot\vec b}}{\vec q^2 } = \frac{1}{2\pi}\log\left(\frac{L}{b}\right),
\ee
where $K_0(mb)$ is the Bessel-$K$ function of order 0 and $L$ is an IR regulator.
}
\be
\delta(s, b) =\frac{\kappa_0s}{M_{\rm Pl}^{2}}{\bf P}_3^{\rm T}\hat {\cal S}_0(\vec e_1,\vec e_3, i\vec\partial_b)\, {\bf P}_1
 \frac{1}{2\pi}\log\left(\frac{L}{b}\right)+\frac{\kappa_ms}{M_{\rm Pl}^{2}}{\bf P}_3^{\rm T}\hat {\cal S}_m(\vec e_1,\vec e_3, i\vec\partial_b)\, {\bf P}_1
\frac{1}{2\pi}K_0(mb).
\label{eq:shockwaveamplitude}
\ee
Here the term proportional to $\kappa_0$ comes from the diagrams with a massless spin-2 particle on the internal line and the term proportional to $\kappa_m$ comes from the exchange diagrams involving a massive spin-2 particle.

The operator $\hat{\cal S}_0$ is given explicitly by

\bea
\tiny
\nonumber
\hat{\cal S}_0=\left(\begin{array}{cccc}
\frac{b_1}{2}& \frac{{\cal C}^0_{SV}}{m}e^i_1\partial_{b^i} & \frac{{\cal C}^0_{ST}}{m^2}e_1^{ij}\partial^2_{b^{ij}} &\frac{-c_2}{\sqrt 6m^2}e^{ij}_1\partial^2_{b^{ij}}  \\
\frac{{\cal C}^0_{SV}}{m}e^i_{3}\partial_{b^i} &\frac{b_1}{2}\vec e_1\cdot\vec e_{3}+\frac{{\cal C}^0_{VV} }{m^2}\vec e_1\cdot\vec\partial_b \vec e_{3}\cdot\vec\partial_b&\frac{{\cal C}_{TV}^0 }{m}e_1^{ij}e_{3}^i\partial_{b^j}+\frac{b_5}{2\sqrt2m^3}e_1^{ij}e_{3}^k\partial^3_{b^{ijk}}& \frac{\sqrt 2c_2}{ m^3}e_1^{ij}e_3^k \partial^3_{b^{ijk}}\\
\frac{{\cal C}^0_{ST}}{m^2}e_{3}^{ij}\partial^2_{b^{ij}} &\frac{{\cal C}_{TV}^0 }{m}e_{3}^{ij}e_1^i\partial_{b^j}+\frac{b_5}{2\sqrt 2m^3}e_{3}^{ij}e_1^k\partial^3_{b^{ijk}} & {b_1\over 2} e_1^{ij} e_{3}^{ij}+\frac{b_5}{2m^2}e_1^{ij}e_{3}^{jk}\partial^2_{b^{ik}}+\frac{b_6}{2m^4}e_1^{ij}e_{3}^{kl}\partial^4_{b^{ijkl}} &\frac{2c_2}{m^4}e_1^{ij}e_{3}^{kl}\partial^4_{b^{ijkl}}\\
\frac{-c_2}{\sqrt 6m^2}e^{ij}_3\partial^2_{b^{ij}} &\frac{\sqrt 2c_2}{ m^3}e_3^{ij}e_1^k \partial^3_{b^{ijk}}&\frac{2c_2}{m^4}e_1^{ij}e_{3}^{kl}\partial^4_{b^{ijkl}}&\frac{d_1}{2}e_1^{ij} e_{3}^{ij}+\frac{d_3}{2m^4}e_1^{ij}e_{3}^{kl}\partial^4_{b^{ijkl}}
\end{array}
\right),
\normalsize \\ \label{eq:genpolmatrixmassless}
\eea

where we have defined $\partial^n_{b^{i_1\cdots i_n}} \equiv \partial_{b^{i_1}}\cdots \partial_{b^{i_n}}$, and the various constants that appear in the matrix are
\begin{subequations}
\begin{align}
{\cal C}^0_{SV} &= -\frac{\sqrt3}{4}(2b_1-b_2), & {\cal C}_{ST}^0 \, &=  -\frac{1}{2\sqrt 6}\left(2 b_1-2b_2+2b_3+b_5\right)\, ,\\
{\cal C}^0_{VV}  &= \frac{2b_1-2b_2+2b_3+b_5}{4} \, ,& {\cal C}^0_{TV}  &=  \frac{2b_1-b_2}{2\sqrt 2}\, .
\end{align}
\end{subequations}
The operator $\hat{\cal S}_m$ is given by
\be
\tiny
\nonumber
\hat{\cal S}_m=\left(\begin{array}{cccc}
{\cal C}^m_{SS} & \frac{{\cal C}^m_{SV}}{m}e^i_1\partial_{b^i} & \frac{{\cal C}^m_{ST}}{m^2}e_1^{ij}\partial^2_{b^{ij}} & \frac{{\cal C}^m_{SQ}}{m^2}e^{ij}_1\partial^2_{b^{ij}}  \\
\frac{{\cal C}^m_{SV}}{m}e^i_{3}\partial_{b^i} &{\cal C}^m_{VV_1}  \vec e_1\cdot\vec e_{3}+\frac{{\cal C}^m_{VV_2} }{m^2}\vec e_1\cdot\vec\partial_b \vec e_{3}\cdot\vec\partial_b&\frac{{\cal C}_{TV}^m }{m}e_1^{ij}e_{3}^i\partial_{b^j}+\frac{a_5}{2\sqrt 2m^3}e_1^{ij}e_{3}^k\partial^3_{b^{ijk}}& \frac{{\cal C}_{QV_1}}{m}e_1^{ij}e_3^j\partial_{b^j}+\frac{{\cal C}_{QV_2}}{m^3}e_1^{ij}e_3^k \partial^3_{b^{ijk}}\\
\frac{{\cal C}^m_{ST}}{m^2}e_{3}^{ij}\partial^2_{b^{ij}} &\frac{{\cal C}_{TV}^m }{m}e_{3}^{ij}e_1^i\partial_{b^j}+\frac{a_5}{2\sqrt 2m^3}e_{3}^{ij}e_1^k\partial^3_{b^{ijk}} & {a_2\over 2} e_1^{ij} e_{3}^{ij}+\frac{a_5}{2m^4}e_1^{ij}e_{3}^{kl}\partial^4_{b^{ijkl}} &\frac{b_3}{2}e^{ij}_1e_3^{ij}+\frac{b_6}{2m^4}e_1^{ij}e_{3}^{kl}\partial^4_{b^{ijkl}}\\
\frac{{\cal C}^m_{SQ}}{m^2}e^{ij}_3\partial^2_{b^{ij}} &\frac{{\cal C}_{QV_1}}{m}e_1^{ij}e_3^j\partial_{b^j}+\frac{{\cal C}_{QV_2}}{m^3}e_1^{ij}e_3^k \partial^3_{b^{ijk}} &\frac{b_3}{2}e^{ij}_3e_1^{ij}+\frac{b_6}{2m^4}e_3^{ij}e_{1}^{kl}\partial^4_{b^{ijkl}}& \frac{c_2}{2}e_3^{ij} e_{1}^{ij}-\frac{2c_2}{m^2}e_3^{ij}e_{1}^{jk}\partial^2_{b^{ik}}+\frac{2c_2}{m^4}e_1^{ij}e_{3}^{kl}\partial^4_{b^{ijkl}}
\end{array}
\right),
\label{eq:genpolmatrixmassive}
\ee
\normalsize
with coefficients
\begin{subequations}
\begin{align}
{\cal C}^m_{SS} &= \frac{1}{16}\left(8 a_1+28 a_2-12 a_3+3 a_5\right)  , & {\cal C}^m_{SV} &= -\frac{1}{8\sqrt 3}\left(4a_1+20a_2-10a_3+3a_5\right),\\
{\cal C}_{ST}^m &=  \frac{1}{4\sqrt 6}\left(4a_3-8a_2-3a_5\right), &{\cal C}^m_{SQ}  &= \frac{1}{2\sqrt 6}\left(2b_2-2b_1-2b_3+2b_5-3b_6\right),\\
{\cal C}^m_{VV_1}   &= \frac{a_1+3a_2-a_3}{4} \, , & {\cal C}^m_{VV_2}  &= \frac{2a_2 -  a_3 + a_5}{4}\, ,\\
{\cal C}^m_{TV}  &=  \frac{2a_2-a_3}{2\sqrt 2} \, , & {\cal C}^m_{QV_1} &=-\frac{b_2-2b_3}{2\sqrt 2 }\, ,\\
{\cal C}^m_{QV_2}&=-\frac{b_5-2b_6}{2\sqrt 2}.
\end{align}
\end{subequations}

The goal now is to place constraints on the $a_i, b_i, c_i$ and $d_i$ coefficients by demanding that the eikonal phase, $\delta$, is positive for all impact parameters and all choices of polarizations for the external particles. The amplitude written in the form~\eqref{eq:shockwaveamplitude} makes manifest that once we have chosen an explicit basis for the polarizations, the amplitude is then a $7 \times 7$ matrix corresponding to the $5+2$ possible polarization states for the incoming and outgoing spin-2 particles. The phase shifts come from diagonalizing this matrix, and it is these eigenvalues which must be non-negative. A convenient way to organize the calculation is to enforce positivity of the phase shift order-by-order in the limit of small impact parameter.

The precise form of the amplitude at each order is not particularly enlightening, so we just quote the parameter constraints. The leading contributions to the phases in the small $b$ limit are of order $b^{-4}$. The phases at this order come from diagonalizing a matrix that depends on the couplings from the six-derivative spin-2 vertices, $a_5$, $b_6$, $d_3$, $c_2$, as well as the scalar couplings, $\kappa_0$ and $\kappa_m$. This matrix has a vanishing trace and hence the sum of eigenvalues vanishes. The eigenvalues can then only have the same sign if they all vanish, which means the matrix itself vanishes since it is symmetric. This gives the simple constraints
\begin{subequations}
\label{eq:scalarconstr1}
\begin{align}
b_6 \kappa_0+a_5 \kappa_m  &= 0,\\
4c_2 \kappa_0+ b_6\kappa_m &=0, \\
d_3\kappa_0+ 4 c_2\kappa_m &=0.
\end{align}
\end{subequations}
Similarly, at order $b^{-3}$ we obtain the constraints
\begin{subequations}
\begin{align}
b_5\kappa_0 + a_5\kappa_m &= 0,  \\
-4c_2\kappa_0 +( b_5-2b_6)\kappa_m &=0 .
\end{align}
\end{subequations}
At order $b^{-2}$ the constraints are
\begin{subequations}
\begin{align}
2(2b_1-2b_2+2b_3+b_5)\kappa_0 +( 8 a_2-4a_3+3a_5)\kappa_m & = 0,  \\
(2b_1-2b_2+2b_3+b_5)\kappa_0+( 2a_2-a_3+a_5)\kappa_m &= 0, \\
2 c_2 \kappa_0+ (2 b_1-2b_2+2b_3-2b_5+3b_6)\kappa_m &=0, \\
a_5\kappa_m =b_6\kappa_m =c_2 \kappa_m &=0.
\end{align}
\end{subequations}
 Lastly, at order $b^{-1}$ we get the constraints
\begin{subequations}
\label{eq:scalarconstr2}
\begin{align}
6(2b_1-b_2)\kappa_0 + (4a_1+20a_2-10a_3+3a_5)\kappa_m &= 0,  \\
(-2b_1+b_2)\kappa_0+  (-2a_2+a_3-a_5)\kappa_m &=0,\\
(-2b_1+b_2)\kappa_0+  (-2a_2+a_3)\kappa_m &=0, \\
(b_2-2b_3)\kappa_m = (b_2-2b_3+b_5-2b_6)\kappa_m &= 0.
\end{align}
\end{subequations}
There are various solutions to these sets of constraints depending on whether or not the scalar couplings, $\kappa_0$ and $\kappa_m$, vanish:
\begin{enumerate}[label=\textbf{\arabic*}.]
\item The trivial case $\kappa_0=\kappa_m=0$ does not constrain the spin-2 couplings since there is no scattering.\footnote{Of course, there will still be constraints on the various couplings coming from other processes.} 

\vspace{.25cm}
\item The case $\kappa_0=0$ with $\kappa_m \neq 0$---which means the scalar does not couple to the massless spin-2 particle---has the solution
\begin{subequations}
\begin{align} 
\label{eq:k1zero}
a_1&=a_5=b_5=b_6=c_2=0, \\
a_3&=2a_2,\\
2 b_1&=2 b_3=b_2.
\end{align}
\end{subequations}
In this case the $S$-matrix equivalence principle requires that the massive spin-2 particle also does not couple to massless spin-2 particle (since we have assumed $\kappa_m \neq 0$ so the scalar {\it does} couple to the massive spin-2 particle), so we must additionally have $b_i=0$. The resulting constraints are then equivalent to those found in~\cite{Hinterbichler:2017qyt} for an isolated massive spin-2 particle.

\vspace{.25cm}
\item The case $\kappa_m=0$ and $\kappa_0 \neq 0$---which means the scalar does not couple to the massive spin-2 particle---has the solution
\begin{subequations}
\begin{align}
\label{eq:k2zero}
b_5&=b_6=0=c_2=d_3=0, \\
 2b_1&= 2b_3=b_2.
\end{align}
\end{subequations}
The $a_i$ couplings are unconstrained by this process since the vertex $\mathcal{V}_a$ does not contribute to the amplitude in this case. Substituting these constraints into the general expression for the vertices in the theory, we find that there are two allowed couplings: one which couples 3 massless particles, and one which couples two massive particles with a massless spin-2 particle. Interestingly, both surviving constrained vertices take the Einstein--Hilbert form,
\begin{align} 
\mathcal{V}_b & = \frac{i b_1}{M_{\rm Pl}} \left(z_{2 3} zp_{1 2} + z_{1 3} zp_{2 3} + z_{12} zp_{3 1}\right)^2, \label{VbEH}  \\
\mathcal{V}_d & = \frac{i d_1}{M_{\rm Pl}} \left(z_{2 3} zp_{1 2} + z_{1 3} zp_{2 3} + z_{12} zp_{3 1}\right)^2  .\label{VdEH}
\end{align} 
The $S$-matrix equivalence principle requires that $b_1=d_1=\kappa_0$.  We can further fix $d_1=2$ by canonically normalizing $M_{\rm Pl}$, leaving no free parameters beyond $M_{\rm Pl}$.

\vspace{.25cm}
\item The last case is where we have both $\kappa_0 \neq 0$ and $\kappa_m \neq 0$, which has the solution
\begin{subequations}
 \label{eq:genconstraints}
\begin{align}
a_1&=a_5=b_5=b_6=c_2=d_3=0,\\
 a_3 &= 2 a_2,\\
  2 b_1&=2 b_3=b_2.
\end{align}
\end{subequations}
The difference with the previous case is that there is now also a coupling between 3 massive spin-2 particles.
The surviving vertices all take the Einstein--Hilbert form:~\eqref{VbEH},~\eqref{VdEH}, and
\begin{align} 
\mathcal{V}_a & = \frac{i a_2}{M_{\rm Pl}} \left(z_{2 3} zp_{1 2} + z_{1 3} zp_{2 3} + z_{12} zp_{3 1}\right)^2. 
\end{align} 
Again, the $S$-matrix equivalence principle requires $b_1=d_1=\kappa_0$, and we can further fix $d_1=2$ by canonically normalizing $M_{\rm Pl}$.  This leaves two free parameters: $a_2$ and $\kappa_m$.
\end{enumerate}

\begin{figure}
\begin{center}
\epsfig{file=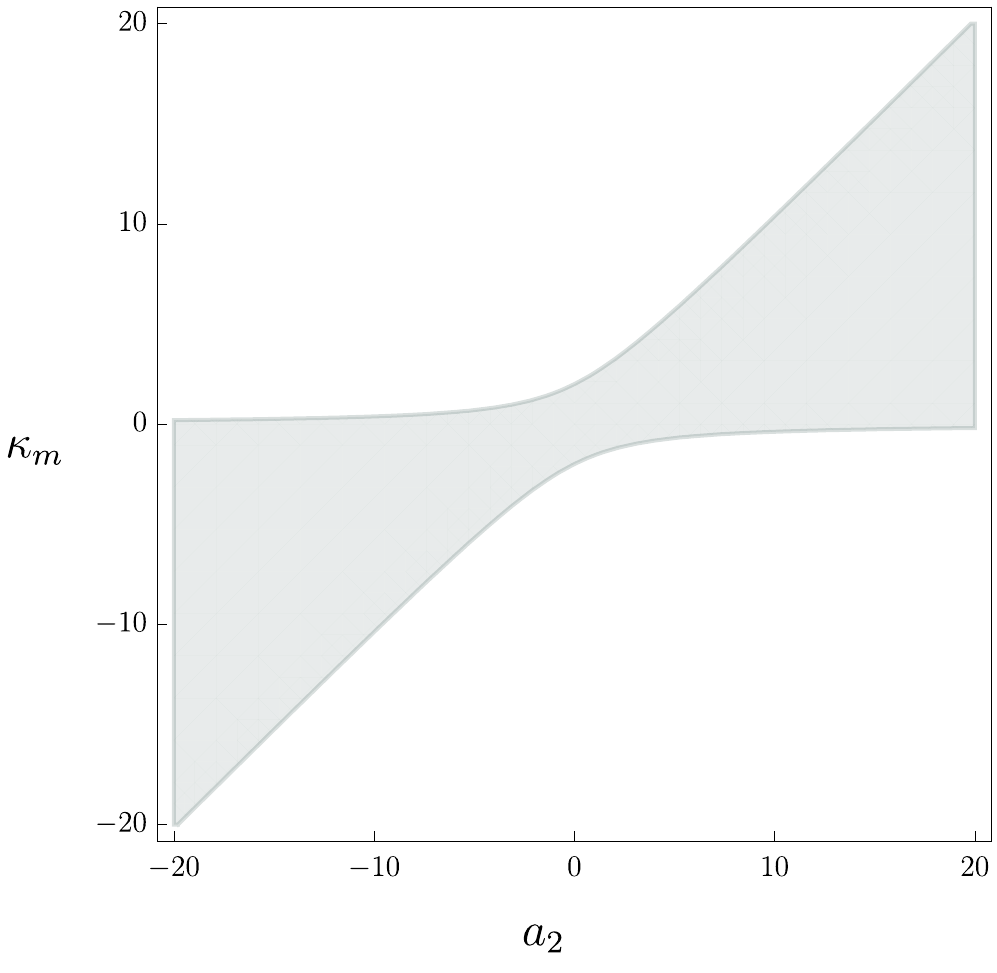,scale=.8}
\caption{\small Plot of the allowed parameters that satisfy the inequalities~\eqref{constraint}.}
\label{fig:ineqplot}
\end{center}
\end{figure}

After solving the constraints~\eqref{eq:scalarconstr1}--\eqref{eq:scalarconstr2}, we still have to check that the remaining phase shifts are positive in the non-trivial cases. 
\begin{itemize}
\item
For the case $\kappa_0=0$, $\kappa_m \neq 0$, the remaining phase shifts are 
\begin{subequations}
\begin{align}
\label{phasek1zero}
\delta_{S,V} &= a_2\frac{\kappa_m s}{32 \pi M_{\rm Pl}^2} K_0(b m),\\
\label{eq:phasek1zerotensors}
\delta_T &=  \left(a_2 \pm \sqrt{a_2^2 + 4 b_1^2}\right)\frac{\kappa_m s}{32 \pi M_{\rm Pl}^2} K_0(b m),
\end{align}
\end{subequations}
where $\delta_{S,V}$ is the phase shift for both the scalar and vector modes, while $\delta_T$ is the phase shift for the tensor modes. 
Since $b_1^2 \geq 0$, some of these are negative unless $b_1=0$. This is the same constraint that the $S$-matrix equivalence principle would impose. The remaining phase shifts are then all non-negative for $\kappa_m a_2 \geq 0$, as found in~\cite{Hinterbichler:2017qyt}.

\item 
In the case $\kappa_m=0$, $\kappa_0 \neq 0$, the final phase shifts are
\begin{align} \label{phasef2zero}
\delta  & = \frac{ \kappa_0 s}{16 \pi M_{\rm Pl}^2} \log \left(\frac{L}{b}\right) \times \left( b_1 \text{ or } d_1 \right),
\end{align}
for all polarizations, where $L$ is an IR regulator. These are positive for $\kappa_0 b_1>0$ and $\kappa_0 d_1>0$, as guaranteed by the $S$-matrix equivalence principle constraint $b_1=d_1=\kappa_0$.

\item
For the final case $\kappa_0 \neq 0$ and $\kappa_m \neq 0$, the general form of the phase-shifts is a bit complicated but after imposing the $S$-matrix equivalence principle constraints, $b_1=d_1=\kappa_0$, we get
\begin{subequations}
 \label{eq:phase1}
\begin{align}
\delta_{S,V} &= \frac{s}{32 \pi M_{\rm Pl}^2} \left[ 2  b_1^2 \log \left( \frac{L}{b} \right) +a_2\kappa_m K_0(b m)\right] ,\\
\delta_T &= \frac{s}{32 \pi M_{\rm Pl}^2} \left[ 2  b_1^2 \log \left( \frac{L}{b} \right) + \left(a_2\pm \sqrt{a_2^2 + 4 b_1^2}\right)\kappa_m K_0(b m) \right] .
\end{align}
\end{subequations}
The $\delta_{S,V}$ phase-shifts comes from the scalar and vector modes, which are each diagonal.  The $\delta_T$ phase-shifts come from the tensor modes of the massive and massless spin-2 particles, which are mixed.
The terms in square brackets have the general form
\be
A K_0(b m) - 2 b_1^2 \log (bm)+2 b_1^2 \log(Lm),
\ee
where $A$ is some combination of coupling constants.
In the limit $b m\rightarrow 0$ this becomes
\be
(-A - 2 b_1^2) \log (bm)+\log2-\gamma_{\rm E}+2 b_1^2 \log(Lm) +\ldots,
\ee
which divergences like $ \log (bm)$, {\it i.e.}, a negative phase shift at small $b$, unless $A \geq -2 b_1^2$. The total phase shift still has the wrong sign for large $bm$, but the difference between this and the GR result can be made arbitrarily small by increasing $L m$. 
The condition $A \geq -2 b_1^2$ in \eqref{eq:phase1} gives
\be
\label{constraint}
2 b_1^2+ \kappa_m \left(a_2\pm \sqrt{a_2^2 + 4 b_1^2}\right) \geq  0.
\ee
These conditions are automatically satisfied when we turn off the scalar coupling to the massive spin-2 particle, $\kappa_m \rightarrow 0$, which is consistent with \eqref{phasef2zero}. We can use the freedom to rescale the Planck mass to fix $b_1 = 2$, after which the remaining inequalities define an allowed two-dimensional region, which we plot in Figure~\ref{fig:ineqplot}.
In Section \ref{sec:shockwave} we will discuss how this constrains matter couplings in ghost-free bi-gravity. 
\end{itemize}

\subsection{Spin-2--Spin-2 eikonal scattering}
\label{sec:spin2spin2}

We now consider the eikonal scattering between massive and massless spin-2 particles where all external legs have spin 2 in $D=4$. There are $2^5=32$ different diagrams that contribute to this amplitude, since each of the four external legs and the internal line can be either massive or massless. As a result the computation is substantially more intricate, but the final result can be phrased fairly simply in terms of the ingredients we have already introduced:
\begin{align}
\nonumber
\delta(s, b) &= \frac{s}{M_{\rm Pl}^{2}}{\bf P}_{3,4}^{\rm T}\hat {\cal S}_0(\vec e_1,\vec e_3, i\vec\partial_b)\otimes \hat {\cal S}_0(\vec e_2,\vec e_4, -i\vec\partial_b)\, {\bf P}_{1,2}
 \frac{1}{2\pi}\log\left(\frac{L}{b}\right)\\
 &+\frac{s}{M_{\rm Pl}^{2}}{\bf P}_{3,4}^{\rm T}\hat {\cal S}_m(\vec e_1,\vec e_3, i\vec\partial_b)\otimes \hat {\cal S}_m(\vec e_2,\vec e_4, -i\vec\partial_b)\, {\bf P}_{1,2}
\frac{1}{2\pi}K_0(mb).
\label{eq:generalamplitude}
\end{align}
Here $\hat{\cal S}_0$ and $\hat{\cal S}_m$ are the same matrices defined in Section~\ref{eq:d4scalarspin2} and the polarization information is now carried by ${\bf P}_{a,b} \equiv {\bf P}_a\otimes{\bf P}_b$ which is a tensor product of the vectors of coefficients for the top and bottom vertices. The fact that our result can be written in the form \eqref{eq:generalamplitude} is a nontrivial check of our calculation.  

Going through a similar procedure to that described in Section~\ref{eq:d4scalarspin2}, we find that the leading contributions in the limit of small impact parameter start at order $b^{-8}$. Constraining the phase shifts to 
be positive all the way through to order $b^{-1}$ reproduces the constraints~\eqref{eq:genconstraints} that we found when both particles coupled to 
matter. This is not unexpected since the spin-2--scalar process should be captured by the more general process considered here after averaging over the polarizations of the lower half of the diagram.

If we impose the $S$-matrix equivalence principle constraint that $d_1=b_1$, then the remaining phase-shifts are given by
\begin{align}
\delta & =  \frac{s }{32 \pi M_{\rm Pl}^2} \left[2 b_1^2 \log\left( \frac{L}{b} \right) + A K_0(b m) \right],
\end{align}
where $A$ is one of the following combinations of the coupling constants:
\be
A=
\left\{
\begin{array}{l}
-2 b_1^2, \\ \frac{a_2^2}{2}, \\ \frac{1}{2}\left( a_2^2 \pm a_2 \sqrt{a_2^2+4 b_1^2}\right), \\ a_2^2+2 b_1^2 \pm a_2 \sqrt{a_2^2+4 b_1^2}.
\end{array}
\right.
\ee
These come from diagonalizing a $49 \times 49$ matrix with lots of off-diagonal terms. As discussed earlier, we impose the requirement $A\geq -2 b_1^2$. The first phase shift saturates this constraint, so this theory is only marginally consistent with asymptotic subluminality. All of the other phase shifts automatically satisfy this constraint as well, so we see that there are no further constraints beyond~\eqref{eq:genconstraints}. This is similar to what happens for massless~\cite{Camanho:2014apa} or massive~\cite{Camanho:2016opx,Hinterbichler:2017qyt} spin-2 particles in isolation: the shockwave amplitude happens to capture all the parameter constraints required for the theory to be asymptotically subluminal.

To summarize, we find that it is possible to couple $D=4$ Einstein gravity to a massive spin-2 particle and still satisfy the constraints that follow from forbidding asymptotic superluminality. This requires the vanishing of the cubic vertices involving two massless gravitons and a massive spin-2 particle, $\mathcal{V}_c=0$, as already shown in~\cite{Camanho:2014apa}. Moreover, the remaining interaction vertices must take the Einstein--Hilbert form,
\begin{align} 
\mathcal{V}_a & = \frac{i a_2}{M_{\rm Pl}} \left(z_{2 3} zp_{1 2} + z_{1 3} zp_{2 3} + z_{12} zp_{3 1}\right)^2, \\
\mathcal{V}_b & = \frac{i b_1}{M_{\rm Pl}} \left(z_{2 3} zp_{1 2} + z_{1 3} zp_{2 3} + z_{12} zp_{3 1}\right)^2, \\
\mathcal{V}_d & = \frac{i b_1}{M_{\rm Pl}} \left(z_{2 3} zp_{1 2} + z_{1 3} zp_{2 3} + z_{12} zp_{3 1}\right)^2,
\end{align} 
where we have used the $S$-matrix equivalence principle to match the coefficients of $\mathcal{V}_b$ and $\mathcal{V}_d$. The coupling of the massive spin-2 particle to a scalar is further constrained by the conditions \eqref{constraint}. We see that there is a recurring theme that the Einstein--Hilbert on-shell cubic amplitude is the only possible structure consistent with positivity of the eikonal phase, so it is tempting to speculate that this observation would continue to be true with the addition of further massive spin-2 particles. That is, it seems that the only couplings allowed between spin-2 particles are  of the Einstein--Hilbert form at cubic order.

\section{Shockwaves and Shapiro Delay in Ghost-Free Bi-Gravity}
\label{sec:shockwave}
In General Relativity, the Shapiro time delay experienced by a particle passing by a massive object can be computed by considering the particle's propagation in a shockwave background given by the Aichelburg--Sexl metric~\cite{Aichelburg:1970dh,Dray:1984ha}.  This delay was shown to be equivalent to that determined by the eikonal scattering calculation in~\cite{Kabat:1992tb}. We would therefore like to match the eikonal scattering results we have obtained to a shockwave calculation.\footnote{The analogous computation in massive gravity has been done in~\cite{Camanho:2016opx,Hinterbichler:2017qyt}.}

Here we consider a particular bimetric theory that can be trusted in the classical nonlinear regime, making the computation particularly convenient, though there should be no obstruction to performing the same computation in an arbitrary theory.  The result will in fact only depend on the cubic vertices of the theory, in agreement with the eikonal calculation.  We consider the analogue of the Shapiro time delay calculated in a shockwave background for this theory, and we find that it identically reproduces the results of the eikonal scattering, as expected.  We also find that, in the formulation of the classical non-linear bi-gravity theory, the constraint on the cubic self-coupling picks out the unique bi-gravity theory with a $\mathbb{Z}_2$ symmetry.

\subsection{Ghost-free bi-gravity}
The non-linear ghost-free bi-gravity theory \cite{Hassan:2011zd} is an extension of dRGT massive gravity \cite{deRham:2010kj} that describes two interacting spin-2 particles---one massive and one massless---and no additional degrees of freedom.  The theory is formulated using two metrics, $g_{\mu\nu}$ and $f_{\mu\nu}$, each with their own Einstein--Hilbert kinetic term and a zero-derivative potential term which mixes the two metrics.  The form of the potential term guarantees the absence of ghosts in the classical theory and in the low energy effective quantum theory. In $D=4$, the ghost-free bi-gravity Lagrangian takes the following form:
\be
\label{L}
{\cal L}= \frac{M_g^2}{2} \sqrt{-g} R(g) +\frac{M_f^2}{2} \sqrt{-f} R(f)-m^2M_*^2 \sqrt{-g} \sum_{n=0}^4 \beta_n S_n (\sqrt{g^{-1}f}) \, .
\ee
The $S_n$ are the $n$-th elementary symmetric polynomials of the eigenvalues of the matrix square root $\mathbb{X}^\mu_{\ \nu}\equiv\sqrt{g^{\mu\lambda}f_{\lambda\nu}}$.  They are given by
\bea
\label{potential}
\begin{array}{l}
S_0 (\mathbb{X})= 1  \, ,  \vspace{.1cm} \\
S_1(\mathbb{X})= [\mathbb{X}]  \, ,  \vspace{.1cm} \\
S_2(\mathbb{X})= \tfrac{1}{2}([\mathbb{X}]^2-[\mathbb{X}^2]) \, ,  \vspace{.1cm} \\
S_3(\mathbb{X})= \tfrac{1}{3!}([\mathbb{X}]^3-3[\mathbb{X}][\mathbb{X}^2]+2[\mathbb{X}^3]) \, ,  \vspace{.1cm} \\
S_4(\mathbb{X})=\tfrac{1}{4!}([\mathbb{X}]^4-6[\mathbb{X}]^2[\mathbb{X}^2]+3[\mathbb{X}^2]^2   
+8[\mathbb{X}][\mathbb{X}^3]-6[\mathbb{X}^4])\, ,
\end{array}
\eea
where the square brackets denote the trace of the enclosed matrix.

There are two mass scales $M_g$ and $M_f$, along with the graviton mass $m$.  The effective mass, $M_*$, is defined as
\be
M_* \equiv \left(\frac{1}{M_g^2}+ \frac{1}{M_f^2}\right)^{-1/2} \, .
\ee
The $\beta_n$ are free coefficients. 

If we expand both metrics around the same flat spacetime, then the requirement of no tadpoles for each metric gives two conditions on the $\beta_n$,
\begin{align}
\label{cc}
\beta_0 + 3 \beta_1 + 3 \beta_2 + \beta_3 = 0 \, , \\
\beta_1 + 3 \beta_2 + 3 \beta_3 + \beta_4 = 0 \, .
\end{align}
After imposing this condition, demanding that the parameter $m^2$ is the correctly normalized mass for the massive spin-2 particle requires
\be
\label{mass}
\beta_1 +2\beta_2 +\beta_3 =1 \, .
\ee

These three conditions mean that, of the five $\beta_n$, there are only two free parameters.  These two parameters are often written as $c_3$ and $d_5$ (see, {\it e.g.},~\cite{deRham:2010ik}), where $c_3$ parametrizes cubic interactions in the potential and $d_5$ parametrizes quartic interactions in the potential.  All higher-order interactions in the potential are fixed after specifying $c_3$ and $d_5$.   The $\beta_n$ can be written in terms of $c_3$, $d_5$ as
\begin{subequations} \label{eq:betachange}
\begin{align}
\beta_0 & = 48 d_5+ 24 c_3  -6 , \\
\beta_1 &= - 48 d_5  - 18 c_3 +3,\\
\beta_2 & = 48 d_5+ 12 c_3 -1 ,\\
\beta_3 & =  - 48 d_5-6 c_3,\\
\beta_4 & = 48 d_5.
\end{align}
\end{subequations}

When the $\beta_n$ satisfy $\beta_4 =\beta_0$ and $\beta_3=\beta_1$, the ghost-free bi-gravity theory enjoys a $\mathbb{Z}_2$ symmetry under the interchange of $g_{\mu\nu} \leftrightarrow f_{\mu\nu}$ and $M_g \leftrightarrow M_f$.  From the above expression \eqref{eq:betachange} we see that this occurs precisely when $c_3=1/4$.
The parameter $d_5$ is left unspecified by this requirement.

\subsection{Cubic vertices of ghost-free bi-gravity}
In order to connect to the eikonal computation, we need the cubic vertices that couple the massive and massless particles in bi-gravity. Importantly, 
the $g_{\mu\nu}$ and $f_{\mu\nu}$ metrics are {\it not} mass eigenstates of the bi-gravity theory. If we expand both metrics around the flat background, 
\begin{subequations} \label{eq:metricpert}
\begin{align}
g_{\mu\nu} =\eta_{\mu\nu}+ \frac{2}{M_g} h_{\mu\nu} \, , \\
f_{\mu\nu} =\eta_{\mu\nu}+ \frac{2}{M_f} k_{\mu\nu} \, ,
\end{align}
\end{subequations}
then we find that the mass eigenstates are given by the linear combinations~\cite{Hassan:2011zd}
\begin{subequations} \label{eq:eigenstates}
\begin{align}
\frac{u_{\mu\nu}}{M_*} = \frac{h_{\mu\nu}}{M_f} + \frac{k_{\mu\nu}}{M_g} \, , \\
\frac{v_{\mu\nu}}{M_*} = \frac{h_{\mu\nu}}{M_g} - \frac{k_{\mu\nu}}{M_f} \, ,
\end{align}
\end{subequations} 
so that the quadratic Lagrangian takes the form\footnote{Here $\hat{\cal E}^{\mu\nu\alpha\beta}$ is the Lichnerowicz operator, which acts on symmetric tensors, $X_{\mu\nu}$, as
\be
\hat {\cal E}^{\mu\nu\alpha\beta}X_{\alpha\beta} = -\frac{1}{2} \left(\square X^{\mu\nu} -2 \partial^{(\mu}\partial_{\rho}X^{\nu)\rho} +\partial^\mu\partial^\nu X-\eta^{\mu\nu}\square X + \eta^{\mu\nu} \partial_\rho\partial_\lambda X^{\rho\lambda}
\right).
\ee
}
\be
{\cal L}^{(2)}  = u_{\mu\nu} \hat {\cal E}^{\mu\nu\alpha\beta}u_{\alpha\beta}+ v_{\mu\nu} \hat {\cal E}^{\mu\nu\alpha\beta}v_{\alpha\beta}  -\frac{m^2}{2} \left(v_{\mu\nu}v^{\mu\nu}-v^\mu_{~\mu}v^\nu_{~\nu} \right) \, .
\ee
The fluctuation $u_{\mu\nu}$ is the {massless} spin-2 field and $v_{\mu\nu}$ is the {massive} spin-2 field.  We recover dRGT massive gravity in the limit that $M_f \rightarrow \infty$.  This gives $M_* \rightarrow M_g$, $u_{\mu\nu} \rightarrow  k_{\mu\nu}$ and $v_{\mu\nu} \rightarrow  h_{\mu\nu}$.  The massless $f_{\mu\nu}$ metric decouples and we are left with the massive $g_{\mu\nu}$ metric.

If we expand the ghost-free bi-gravity Lagrangian \eqref{L} to cubic order and use the mass eigenstates \eqref{eq:eigenstates} we find the following four terms in the on-shell cubic Lagrangian: 
\begin{align}
{\cal L}_{\rm on-shell}^{(3)}\,  =\,&\frac{M_*}{M_g M_f} {\cal L}_{EH}^{(3)}(u)
+ \left(\frac{M_*}{M_g^2}-\frac{M_*}{M_f^2}\right) {\cal L}_{EH}^{(3)}(v)
+ \frac{M_*}{M_g M_f} {\cal L}^{(3)}(u\nabla v \nabla v) \nonumber \\
&-m^2\left[\frac{M_*}{M_g^2}\left(2 c_3-\frac{3}{2}\right) + \frac{M_*}{M_f^2}\left(2 c_3+\frac{1}{2}\right) \right]v_{\mu\nu}^3 \, .
\end{align}
Both the massless $u$ field and massive $v$ field have an Einstein--Hilbert cubic vertex with different coefficients; there is also a two-derivative term containing one massless and two massive spin-2 particles. The coefficients of all three of these interactions are completely fixed once the mass scales have been chosen. Finally, there is also a cubic potential for the massive spin-2 particle, with a free coefficient parameterized by $c_3$.
The corresponding on-shell cubic vertices are given by
\begin{align}
\mathcal{V}_a & = 2 i M_{*} \left( \frac{1}{M_g^2}-\frac{1}{M_f^2} \right) \left(z_{2 3} zp_{1 2} + z_{1 3} zp_{2 3} + z_{12} zp_{3 1}\right)^2+\frac{3 i m^2(1-4 c_3)}{M_*}z_{12}z_{13}z_{23}, \\
\mathcal{V}_b & = \frac{2 i M_{*}}{M_g M_f}  \left(z_{2 3} zp_{1 2} + z_{1 3} zp_{2 3} + z_{12} zp_{3 1}\right)^2, \\
\mathcal{V}_d & = \frac{2 i M_{*}}{M_g M_f}  \left(z_{2 3} zp_{1 2} + z_{1 3} zp_{2 3} + z_{12} zp_{3 1}\right)^2.
\end{align}
The mass scale suppressing the massless cubic vertex, ${\cal V}_d$, is $M_f M_g/M_* = \sqrt{M_f^2+M_g^2}$ so this defines $M_{\rm Pl}$ for this theory. Comparing to the vertices in Section \ref{sec:cubic}, we then see that ghost-free bi-gravity corresponds to
\begin{align} \label{eq:bgcouplings}
a_1 &= 3\frac{M_f M_g}{M_*^2} (1-4c_3) , \quad a_2=  2\! \left( \frac{M_f }{M_g}-\frac{M_g}{M_f} \right) , \quad a_3 = 2\, a_2  , \quad a_5=0,\\b_1& =d_1=b_3=2 , \quad b_2 =4  , \quad c_2=d_3=b_5=b_6=0  .
\end{align}
The original bi-gravity action is invariant under a diffeomorphism which transforms both metrics as tensors, and so we automatically have $b_1=d_1$ as mandated by the $S$-matrix equivalence principle. 

In the massive gravity limit $M_f \rightarrow \infty$ we are left with just the vertex 
\be
\mathcal{V}_a  =  \frac{2 i }{M_g} \left(z_{2 3} zp_{1 2} + z_{1 3} zp_{2 3} + z_{12} zp_{3 1}\right)^2+\frac{3 i m^2(1-4 c_3)}{M_g}z_{12}z_{13}z_{23}, 
\ee
which is the dRGT cubic vertex~\cite{Hinterbichler:2017qyt}.

\subsection{Background solutions}
We now look for shockwave solutions of the ghost-free bi-gravity theory~\eqref{L}. We use light-cone coordinates~\eqref{eq:lconecoords}, and look for Aichelburg--Sexl type solutions of the following form:
\begin{align}
\bar{g}_{\mu\nu} = \eta_{\mu\nu}+ G(x^{\scriptscriptstyle +},\vec x)\, l_\mu l_\nu \, , \\
\bar{f}_{\mu\nu} = \eta_{\mu\nu}+ F(x^{\scriptscriptstyle +}, \vec x)\, l_\mu l_\nu \, ,
\end{align}
where $l^\mu = (1,0,\vec 0)$.  Substituting this ansatz into the equations of motion for $g$ and $f$ that follow from \eqref{L}, we find
\begin{align}
\label{eq:aseqs1}
-\frac{M_g^2}{2} \nabla^2 G(x^{\scriptscriptstyle +},\vec x)+\frac{m^2M_*^2}{2}\left[G(x^{\scriptscriptstyle +},\vec x)-F(x^{\scriptscriptstyle +},\vec x)\right]=T_{{\scriptscriptstyle ++}}^{(g)} \, ,\\
\label{eq:aseqs2}
-\frac{M_f^2}{2} \nabla^2 F(x^{\scriptscriptstyle +},\vec x)+\frac{m^2M_*^2}{2}\left[F(x^{\scriptscriptstyle +},\vec x)-G(x^{\scriptscriptstyle +},\vec x)\right]=T_{{\scriptscriptstyle ++}}^{(f)} \, ,
\end{align}
where $ \nabla^2 \equiv \partial_y^2 +\partial_z^2$.  Notably, the $\beta_n$ disappear from these equations as we would expect for an Aichelburg--Sexl type metric, which typically only depends on the linear structure of the theory.

The stress tensor sources that we consider are
\begin{align}
T_{{\scriptscriptstyle ++}}^{(g)} =\frac{M_g}{M_f}   |p^{\scriptscriptstyle -}| \, A_0(x^{\scriptscriptstyle +})\, \delta(\vec x)+|p^{\scriptscriptstyle -}|\,  A_m(x^+)\, \delta(\vec x) \, , \\
T_{{\scriptscriptstyle ++}}^{(f)} =  \frac{M_f}{M_g} |p^{\scriptscriptstyle -}| \, A_0(x^{\scriptscriptstyle +})\, \delta(\vec x)- |p^{\scriptscriptstyle -}|\,  A_m(x^{\scriptscriptstyle +})\, \delta(\vec x) \,,
\end{align}
which in the general case describe shocks propagating in the $x^{\scriptscriptstyle -}$ direction coupled to both the massive and massless eigenstates with radial profiles given by the general functions $A_0(x^{\scriptscriptstyle +})$ and $A_m(x^{\scriptscriptstyle +})$.
From \eqref{eq:eigenstates}, the coupling of the mass eigenstates to the energy-momentum tensor is given by
\be
\frac{2}{M_g}h^{{\scriptscriptstyle ++}}\, T_{{\scriptscriptstyle ++}}^{(g)} + \frac{2}{M_f}k^{{\scriptscriptstyle ++}}\, T_{{\scriptscriptstyle ++}}^{(f)}  = \frac{2 }{M_*}   u^{{\scriptscriptstyle ++}}\,  |p^{\scriptscriptstyle -}| A_0(x^{\scriptscriptstyle +})\, \delta(\vec x)+  \frac{2 }{M_*}  v^{{\scriptscriptstyle ++}}\,  |p^{\scriptscriptstyle -}| A_m(x^{\scriptscriptstyle +})\, \delta(\vec x)\,.
\ee
We see that the function $A_0(x^{\scriptscriptstyle +})$ sources the massless spin-2 particle $u_{\mu\nu}$ while $A_m(x^{\scriptscriptstyle +})$ sources the massive spin-2 particle $v_{\mu\nu}$. We consider a point source energy-momentum tensor and define the constants $\kappa_0$ and $\kappa_m$ by
\be
\label{eq:As}
A_0(x^{\scriptscriptstyle +}) = \kappa_0  \frac{M_*^2}{2 M_f M_g} \delta(x^{\scriptscriptstyle +}), \quad A_m(x^{\scriptscriptstyle +}) = \kappa_m  \frac{M_*^2}{2 M_f M_g} \delta(x^{\scriptscriptstyle +}),
\ee
which correspond to the scalar couplings defined in Section \ref{sec:scalarspin2}.
The $S$-matrix equivalence principle requires that $\kappa_0=2$.  This choice ensures that matter is canonically coupled to the Minkowski metric in the limit of zero metric fluctuations for both the $f$ and $g$ metrics.

There are two independent solutions to the equations~\eqref{eq:aseqs1} and~\eqref{eq:aseqs2}, which can be expressed as
\begin{align}
G(x^+,\vec x) = A_0(x^{\scriptscriptstyle +})\,f_0(\vec x)+\frac{M_*^2}{M_g^2} A_m(x^{\scriptscriptstyle +})\,f_m(\vec x) \, , \\
F(x^+,\vec x) = A_0(x^{\scriptscriptstyle +})\,f_0(\vec x)- \frac{M_*^2}{M_f^2}A_m(x^{\scriptscriptstyle +})\,f_m(\vec x) \, ,
\end{align}
where the functions $f_0(\vec x)$ and $f_m(\vec x)$ satisfy \begin{align}
\label{f1}
\nabla^2 f_0(\vec x) &= -\frac{2}{M_f M_g} |p^{\scriptscriptstyle -}| \delta(\vec x) \, , \\
\label{f2}
(\nabla^2-m^2) f_m(\vec x) &= -\frac{2}{M_*^2} |p^{\scriptscriptstyle -}|  \delta(\vec x) \, .
\end{align}
The solutions to equations \eqref{f1} and \eqref{f2} are given by
\begin{align}
f_0(\vec x) &= \frac{|p^{\scriptscriptstyle -}|}{\pi M_f M_g} \log\left(\frac{L}{b} \right) \, ,\\
f_m(\vec x) &= \frac{|p^{\scriptscriptstyle -}|}{\pi M_*^2} K_0 (mb) \, .
\end{align}
Here we have introduced an IR cutoff, $L$, in the first expression.

\subsection{Fluctuations}

In order to determine the phase shift due to propagation in the shockwave background, we expand the 
two metrics $g_{\mu\nu}$ and $f_{\mu\nu}$ about their respective backgrounds as follows:
\begin{align}
g_{\mu\nu} = \eta_{\mu\nu}+ G(x^{\scriptscriptstyle +},\vec x)\, l_\mu l_\nu+ \frac{2\, }{M_g} h_{\mu\nu} \, , \\
f_{\mu\nu} = \eta_{\mu\nu}+ F(x^{\scriptscriptstyle +},\vec x)\, l_\mu l_\nu+ \frac{2\, }{M_f} k_{\mu\nu} \, .
\end{align}
We expand the equations of motion for $g$ and $f$ to leading order in the fluctuations $h_{\mu\nu}$ and $k_{\mu\nu}$ and look for solutions.  Of the combined 20 components of $h_{\mu\nu}$ and $k_{\mu\nu}$ there will be seven independent propagating degrees of freedom corresponding to the two degrees of freedom of the massless graviton and the five degrees of freedom of the massive spin-2 particle.  We solve for these.  For clarity, we consider the phase shifts due to the two independent background solutions $F, G$ separately. The total phase shift due to both backgrounds will simply be the sum of the two.

\subsubsection{Massless shockwave}

Let us start by setting $\kappa_0 \neq 0, \kappa_m=0$, which corresponds to a shockwave source that only couples directly to the massless spin-2 particle.  After various field redefinitions of $h_{\mu\nu}$ and $k_{\mu\nu}$ we can isolate the five physical modes of the massive spin-2 field, which we denote by $V_I$ where $I=S,V_1,V_2,T_1,T_2$ and the two physical modes of the massless spin-2 field which we denote by $U_I$ where $I=T_1,T_2$. (See \cite{Camanho:2016opx,Hinterbichler:2017qyt} for more details of our method.)  From the full equations of motion for the metric fluctuations, we find the following five equations of motion for the five massive modes, $V_I$:
\be
\partial_{\scriptscriptstyle +} V_I + i |p^{\scriptscriptstyle +}| \frac{m^2+ q^2}{2 (p^{\scriptscriptstyle +})^2} V_I = i \kappa_0\frac{p^{\scriptscriptstyle -} p^{\scriptscriptstyle +} M_*^2}{16 \pi M_f^2 M_g^2} \delta(x^{\scriptscriptstyle +}) \left[4 \log \left(\frac{1}{m b}\right) V_I \right] \, .
\label{eq:masslessVeq}
\ee
For the two massless modes, $U_I$, we find the following:
\be
\partial_{\scriptscriptstyle +} U_I + i |p^+| \frac{q^2}{2 (p^{\scriptscriptstyle +})^2} U_I = i\kappa_0\frac{p^{\scriptscriptstyle -} p^{\scriptscriptstyle +} M_*^2}{16 \pi M_f^2 M_g^2} \delta(x^{\scriptscriptstyle +}) \left[4 \log \left(\frac{1}{m b}\right) U_I \right] \, .
\label{eq:masslessUeq}
\ee
The mixing matrix between the various modes ({\it i.e.}, the right hand side of the above equations) is diagonal, with the eigenvalues for all seven modes given by
\be
\lambda_I = 4 \log \left(\frac{L}{b}\right) \, .
\ee
After diagonalizing the modes, the first order differential equations~\eqref{eq:masslessVeq} and~\eqref{eq:masslessUeq} can be integrated to obtain
\begin{subequations}
\label{eq:integratedshifts}
\begin{align}
V({x^{\scriptscriptstyle +}}) &=  V_I({x^{\scriptscriptstyle +}_0}) e^{-ip^{\scriptscriptstyle +}\int_{{x^{\scriptscriptstyle +}_0}}^{x^{\scriptscriptstyle +}} \rd  {\tilde x^{\scriptscriptstyle +}} \left(\frac{m^2+q^2}{2(p^{\scriptscriptstyle+})^2} -\frac{\kappa_0p^{\scriptscriptstyle -}M_*^2}{16\pi M_f^2M_g^2}\delta( {\tilde x^{\scriptscriptstyle +}}) \lambda_I(b)  \right)},\\
U({x^{\scriptscriptstyle +}}) &=  U_I({x^{\scriptscriptstyle +}_0}) e^{-ip^{\scriptscriptstyle +}\int_{{x^{\scriptscriptstyle +}_0}}^{x^{\scriptscriptstyle +}} \rd  {\tilde x^{\scriptscriptstyle +}} \left(\frac{q^2}{2(p^{\scriptscriptstyle+})^2} -\frac{\kappa_0p^{\scriptscriptstyle -}M_*^2}{16\pi M_f^2M_g^2}\delta( {\tilde x^{\scriptscriptstyle +}}) \lambda_I(b)  \right)}.
\end{align}
\end{subequations}
The phase shifts experienced by the particles have two components, the first is a dispersion effect which is integrated along the trajectory; the second is an anomalous phase shift localized at the shock, which corresponds to the Shapiro time delay. We integrate just across the shock to isolate this second component:
\be
\label{phase}
\delta (s,b) =\kappa_0 \frac{p^{\scriptscriptstyle -} p^{\scriptscriptstyle +} M_*^2}{16\pi M_f^2 M_g^2} \lambda_I (b) 
=\kappa_0 \frac{s }{8 \pi M_{\rm Pl}^2} \log \left(\frac{L}{b}\right)  \, ,
\ee
where $M_{\rm Pl} \equiv M_f M_g/M_*$. This agrees with the corresponding amplitude result \eqref{phasef2zero} when we substitute $b_1=2$. The $S$-matrix equivalence principle sets $\kappa_0=2$ and we see again that all seven modes have a positive phase shift.

\subsubsection{Massive shockwave}

We repeat the same procedure, now setting $\kappa_0= 0, \kappa_m \neq 0$. This corresponds to a shockwave source coupling only to the massive spin-2 particle. For the scalar and vector modes of the massive particle we find the following equations:
\be
\partial_{\scriptscriptstyle +} V_I + i \lvert p^{\scriptscriptstyle +}\rvert \frac{m^2+ q^2}{2 (p^{\scriptscriptstyle +})^2} V_I = i\kappa_m \frac{p^{\scriptscriptstyle -} p^{\scriptscriptstyle +}}{16 \pi M_f M_g} \delta(x^{\scriptscriptstyle +}) M_{IJ} V_J \, ,
\ee
where we restrict to $I=S,V_1,V_2$.  $M_{IJ}$ is a $3 \times 3$ matrix with eigenvalues given by
\begin{align}
\lambda_S  &= \frac{ (5-12 c_3)M_f^2+(1-12c_3)M_g^2 }{M_f^2+M_g^2} K_0(mb) \, , \vspace{0.3cm} \\
\lambda_{V_1,V_2}  &=  \frac{ (13-36 c_3)M_f^2+(5-36c_3)M_g^2 }{2(M_f^2+M_g^2)}  K_0(mb)
 \pm \frac{1- 4 c_3}{2}\sqrt{9 K_0(mb)^2+48 K_1(mb)^2} \, . 
\end{align}
The phase shifts are then given by
\be
\label{phase}
\delta_{S,V} (s,b) =\kappa_m \frac{p^{\scriptscriptstyle -} p^{\scriptscriptstyle +}}{16\pi M_f M_g} \lambda_I (b) 
= \kappa_m \frac{s}{32 \pi M_f M_g}  \lambda_I(b).
\ee
In order for the vector phase shift to be positive at small impact parameter, we must have $c_3 = \tfrac{1}{4}$ and thus $a_1=0$, as found in \eqref{eq:k1zero}.  On this value, the eigenvalues become
\be
\lambda_{S,V}  = 2 \frac{M_f^2-M_g^2 }{M_f^2+M_g^2}  K_0(mb) \, .
\ee
The phase shifts are thus
\be
\delta_{S,V}=\kappa_m \frac{s}{16 \pi M_{\rm Pl}^2} \left( \frac{M_f}{M_g}-\frac{M_g}{M_f}\right) K_0(mb).
\ee
This agrees with the scattering result \eqref{phasek1zero} after substituting $a_2$ as given in \eqref{eq:bgcouplings}.

If we now consider the tensor modes, we find that the equations mix the massless $U_I$ and massive $V_I$ tensor modes:
\begin{align}
\partial_{\scriptscriptstyle +} U_I+ i |p^{\scriptscriptstyle +}| \frac{q^2}{2 (p^+)^2} U_I &= i\kappa_m \frac{p^{\scriptscriptstyle -} p^{\scriptscriptstyle +} M_*^2}{16 \pi M_f^2 M_g^2} \delta(x^{\scriptscriptstyle +}) \left[ 4 K_0(mb) V_I\right] \, , \vspace{0.3cm} \\
\partial_{\scriptscriptstyle +} V_I + i |p^{\scriptscriptstyle +}| \frac{m^2+q^2}{2 (p^+)^2} V_I &= i\kappa_m \frac{p^{\scriptscriptstyle -} p^{\scriptscriptstyle +} M_*^2}{16 \pi M_f^2 M_g^2} \delta(x^{\scriptscriptstyle +}) \left[4 K_0(mb) U_I+ 4\frac{M_f^2-M_g^2}{M_f M_g} K_0(mb) V_I\right] \, .
\end{align}
Here $I=T_1,T_2$ corresponds to the two tensor modes for each of the fields.  As can be seen from the left hand side of the above equations, the two sets of modes obey different dispersion relations---one massless and one massive.  However, this difference becomes negligible when integrating across the shockwave to find the phase difference.  Thus we can simply calculate the eigenvalues of the mixing matrix from the right hand side of the above expression.  We find the following four eigenvalues:
\be
\lambda_{T_{1,2}} = 4 \frac{M_f}{M_g} K_0(mb) \, , \quad
\lambda_{T'_{1,2}} =  - 4 \frac{M_g}{M_f} K_0(mb) \, .
\ee
The phase shifts are thus given by
\be
\delta_{T} =  \kappa_m\, \frac{s}{8 \pi M_{\rm Pl}^2} \frac{M_f}{M_g} K_0(mb)  \, , \quad
\delta_{T'}=  - \kappa_m\, \frac{s}{8 \pi M_{\rm Pl}^2} \frac{M_g}{M_f} K_0(mb) \, .
\ee
These agree with the scattering result~\eqref{eq:phasek1zerotensors} after substituting $b_1$ and $a_2$ as given in \eqref{eq:bgcouplings}. Thus, for the case of $\kappa_0=0$, we see that the phases are only positive when $M_f \rightarrow \infty$,  implying $b_1 \rightarrow 0$, which is also the constraint from the equivalence principle.

\subsubsection{General case}
For the most general case with $\kappa_0\neq 0$ and $\kappa_m \neq 0$, the total phase shifts are simply the sum of the phase shifts found separately above for the two independent shockwave solutions.  In order for the phase shift associated to the vector modes to be positive in this generic case, we again find $c_3 = \frac{1}{4}$.  We can then determine the phase shift for all the modes, subject to this constraint.  One finds
\begin{align}
\delta_{S,V}  &=\frac{s }{8 \pi M_{\rm Pl}^2}  \left[\kappa_0 \log \left(\frac{L}{b}\right) 
+ \frac{\kappa_m}{2} \left( \frac{M_f}{M_g}-\frac{M_g}{M_f}\right) K_0(mb)\right] \, ,\\
\delta_{T} &=\frac{s }{8 \pi M_{\rm Pl}^2}  \left[\kappa_0 \log \left(\frac{L}{b}\right) 
+ \kappa_m\frac{M_f}{M_g} K_0(mb)\right] \, , \\
\delta_{T'} &=\frac{s }{8 \pi M_{\rm Pl}^2}  \left[\kappa_0 \log \left(\frac{L}{b}\right) 
- \kappa_m\frac{M_g}{M_f} K_0(mb)\right] \, .
\end{align}
Positivity of the phase shifts then requires the following two conditions on the free parameters of the bi-gravity theory:
\begin{align}
&\kappa_0+\kappa_m\,\frac{M_f}{M_g}  \geq 0 \, ,\\
&\kappa_0-\kappa_m\,\frac{M_g}{M_f}  \geq 0 \, .
\end{align}
These are the same inequalities as~\eqref{constraint}, but phrased in the variables more natural to bi-gravity.

\begin{figure}[h]
\begin{center}
\epsfig{file=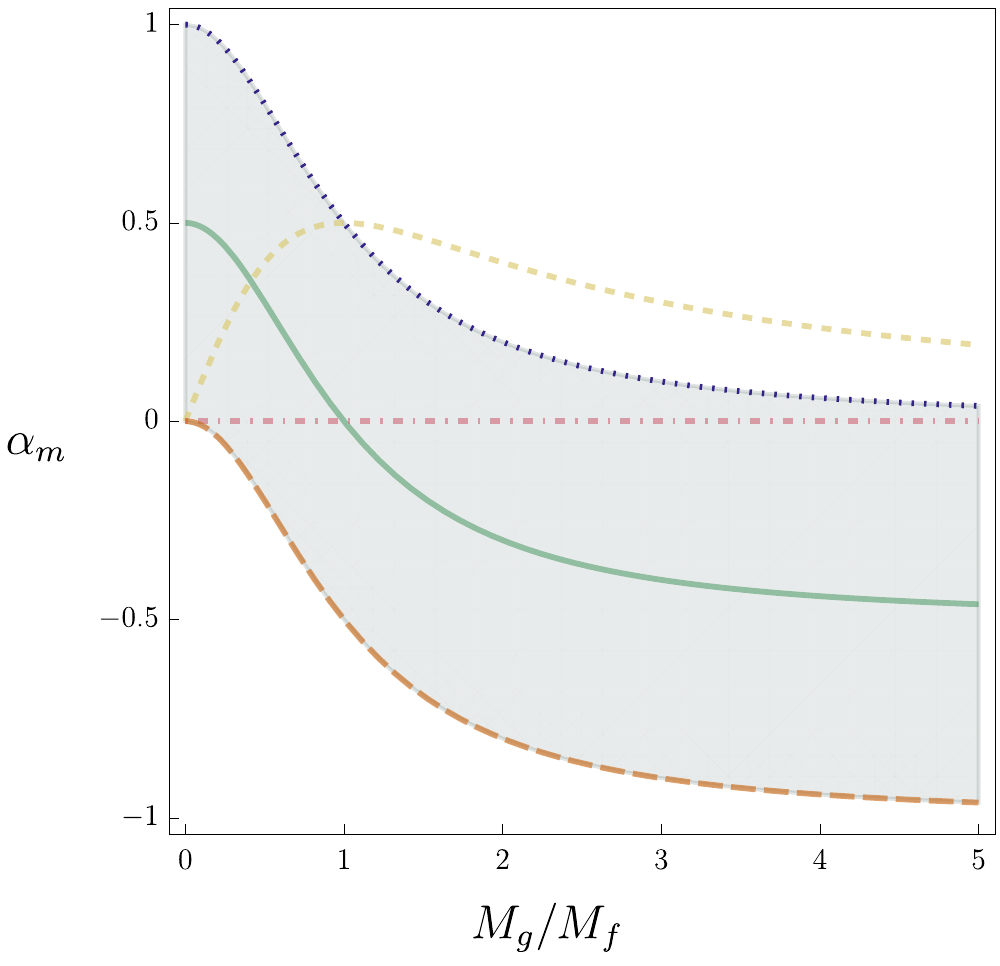,scale=.8}
\caption{\small A plot of the allowed parameter space for the coupling constant $\alpha_m=\kappa_m  \frac{M_*^2}{2 M_f M_g}$  and the ratio $M_g/M_f$.  The shaded region corresponds to the allowed parameter space.  The different curves correspond to different matter couplings: coupling to the $g$ metric only (blue, dotted), to the $f$ metric only (orange, long dash), equally to both metrics (green, solid), to the massless eigenstate only (pink, dot dashed), and equal coupling to both the massive and massless eigenstates (yellow, short dash). The information contained in this plot is equivalent to that in Figure~\ref{fig:ineqplot}, but these variables are the ones that arise naturally in bimetric gravity.}
\label{fig:constraintplot}
\end{center}
\end{figure}

This allowed region is plotted in Figure~\ref{fig:constraintplot}, together with curves corresponding to various special cases of the matter couplings.  For clarity, following the definitions of equation \eqref{eq:As}, we use the couplings
\be
\alpha_0 = \kappa_0  \frac{M_*^2}{2 M_f M_g} \, , \quad \alpha_m = \kappa_m  \frac{M_*^2}{2 M_f M_g} \, .
\ee
The $S$-matrix equivalence principle sets $\kappa_0=2$ and thus $ \alpha_0 = \frac{M_*^2}{M_f M_g}$, leaving an allowed region for values of $\alpha_m$ as a function of $M_g/M_f$.  The dRGT massive gravity limit corresponds to $M_f \rightarrow \infty$ and thus $\alpha_0 \rightarrow 0$.  {\it I.e.}, in this limit the matter sector couples only to the massive eigenstate.

\subsection{Implications for bi-gravity}
As mentioned above, the main constraint that arises from the asymptotic subluminality condition for bi-gravity is that the cubic coupling takes the special value $c_3= {1\over 4}$. Intriguingly, this corresponds to the most general ghost-free bi-gravity theory \eqref{L} with a $\mathbb{Z}_2$ symmetry under the interchange of the $g_{\mu\nu}$ and $f_{\mu\nu}$ metrics.  

In addition, the conditions \eqref{constraint}, found both from the scattering and the shockwave calculations, constrain the matter coupling constant $\kappa_m$ in terms of $M_g/M_f$. The allowed region of parameter space for ghost-free bi-gravity coupled to a single scalar is shown in Figure~\ref{fig:constraintplot}.

Bi-gravity theories in which both metrics couple to the same matter sector have attracted much interest in the literature.  Generically, ``doubly-coupled" matter re-introduces the Boulware--Deser ghost \cite{Boulware:1973my} at scales below the cutoff of the bi-gravity EFT.  However, in \cite{deRham:2014naa} (see also \cite{Noller:2014sta}) a particular double-coupling was introduced of the form
\be
\mathcal{L}_{\phi} = -\frac{1}{2} \sqrt{- g_{\rm eff}} \left( g_{\rm eff}^{\mu \nu} \partial_{\mu} \phi \partial_{\nu} \phi + m_{\phi}^2 \phi^2 \right),
\ee
where 
\be
g^{\rm eff}_{\mu \nu} = \alpha^2 g_{\mu \nu} +2 \alpha \beta g_{\mu \lambda} \mathbb{X}^{\lambda}{}_{\nu}+ \beta^2 f_{\mu \nu}
\ee
and $\mathbb{X}^\mu_{\ \nu}\equiv\sqrt{g^{\mu\lambda}f_{\lambda\nu}}$. With this matter coupling the Boulware-Deser ghost is absent in a certain high-energy limit known as the ``decoupling limit."  This indicates that the scale associated with the ghost is larger than the cutoff $\Lambda_3 = (m^2 M_{Pl})^{1/3}$ of the bi-gravity EFT.  Thus such a matter coupling is consistent within the regime of validity of the EFT. 
Let us make contact between this particular matter coupling and the results of the previous sections.  We can expand the matter Lagrangian around a flat background using \eqref{eq:metricpert} and 
\be
g_{\mu \lambda} \mathbb{X}^{\lambda}{}_{\nu} = \eta_{\mu \nu}+\frac{1}{M_g} h_{\mu \nu}+\frac{1}{M_f} k_{\mu \nu} + \mathcal{O}(h^2, k^2, h k).
\ee
In terms of the mass eigenstates \eqref{eq:eigenstates} this gives the cubic interactions
\be
{\cal L}_\phi\supset-\frac{1}{M_{\rm Pl}} \left[ u^{\mu \nu}+\frac{v^{\mu \nu}}{\alpha+\beta}\left( \alpha \frac{M_f}{M_g}-\beta \frac{M_g}{M_f} \right)  \right] \phi\, \partial_{\mu} \partial_{\nu} \phi,
\ee
where $M_{\rm Pl} = \sqrt{M_f^2 + M_g^2}$ and we have canonically normalized the scalar assuming that $\alpha+\beta \neq 0$. The doubly-coupled matter Lagrangian thus corresponds to the cubic couplings
\be \label{eq:effcoupling}
\kappa_0=2, \quad \kappa_m =  \frac{ 2}{\alpha+\beta} \left(  \alpha \frac{M_f}{M_g} -\beta\frac{M_g}{M_f} \right).
\ee
From this we see that the equivalence principle constraint $\kappa_0=2$ is automatically satisfied, as expected from overall diffeomorphism invariance of the action. The $\kappa_m$ coupling falls inside the region allowed by asymptotic subluminality when $\alpha\beta \geq 0$. The cases of coupling to a single metric, $\alpha=0$ or $\beta=0$, correspond to the top and bottom lines in Figure \ref{fig:constraintplot}, which are on the boundary of the region allowed by asymptotic subluminality. The case $\alpha=\beta=1/2$ gives the solid green curve in Figure \ref{fig:constraintplot}.

\section{Eikonal Scattering in $D>4$}
\label{sec:dneq4amplitudes}
Up to this point, we have focused on physics in $D=4$. Although this is likely the most physically relevant situation, the $D=4$ Gram identity~\eqref{gramidentity} means that some structures vanish identically. We would therefore like to understand if any of these structures can appear in a higher-dimensional theory consistent with asymptotic subluminality. In this section we briefly describe the higher-dimensional version of the eikonal scattering calculation considered in Section~\ref{sec:scalarspin2}.

The main difference in higher dimensions is that all 16 structures can appear. In addition to this, there are now $D-2$ independent vector polarizations and $\frac{(D-2)(D-2)}{2}-1$ tensor polarizations.\footnote{See~\cite{Hinterbichler:2017qyt} for an explicit basis for these in $D$-dimensions.} Aside from this, the computation in $D\neq 4$ is essentially the same as in four dimensions.

Repeating the procedure outlined in Section~\ref{sec:scalarspin2}, we obtain an amplitude of the schematic form~\eqref{eq:shockwaveamplitude}. The various constants appearing in $\hat{\cal S}_0$ and $\hat{\cal S}_m$ are now dimension-dependent. 
Their exact form is not particularly illuminating, so we will not reproduce them here. Upon diagonalizing the amplitude we get a collection of constraints on the couplings. Solving these gives the same solutions as in $D=4$ for the cases $\kappa_0=0, \kappa_m \neq 0$ and $\kappa_m=0, \kappa_0\neq 0 $, namely \eqref{eq:k1zero} and \eqref{eq:k2zero}. The only difference is that the conditions $c_1=d_2=b_4=0$ now follow from positivity of the eikonal phase of the additional vector and tensor polarizations, rather than from dimensionally-dependent identities. The case $\kappa_0 \neq 0, \kappa_m \neq 0$ is different since the higher-dimensional structures can survive. The final solution to the constraints is
\begin{subequations} \label{eq:5dconstraints}
\begin{align}
&a_5=b_6=c_2=d_3=0,\\
&4 a_1=6a_3-12a_2=3 a_4 = -\frac{12 \kappa_0^2 c_1}{\kappa_m^2},\\
& 2b_1 = b_2, \quad b_2-2b_3=b_4=b_5 = \frac{4 \kappa_0 c_1}{\kappa_m}, \quad d_2 = \frac{4 \kappa_m c_1}{\kappa_0}.
\end{align}
\end{subequations}
There is now an extra free coefficient, $c_1$, parameterizing the non-Einstein--Hilbert couplings. However since these additional terms depend on the matter couplings, $\kappa_i$, we should expect them to be set to zero by the full spin-2--spin-2 eikonal calculation. This is because the constraints from the full amplitude must be at least as strong as~\eqref{eq:5dconstraints} but they cannot depend on $\kappa_i$. 

In order to consider pure spin-2 scattering, we square the scalar--spin-2 amplitude as in~\eqref{eq:generalamplitude}. In this case, we find that positivity of the eikonal phase requires, in addition to the constraints~\eqref{eq:5dconstraints}, that $c_1 = 0$, as expected. We therefore see that again all couplings must be of the Einstein--Hilbert form. This is in accordance with the results of~\cite{Camanho:2014apa}, who considered a restricted set of couplings.

\section{Conclusions}

We have derived model independent constraints on the cubic couplings of any effective field theory of a massive spin-2 field coupled to a massless spin-2 field.  These constraints come from demanding positivity of the eikonal phase in $2\rightarrow 2 $ scattering, which corresponds physically to demanding the absence of asymptotic superluminality in the four-particle $S$-matrix.  We find that the vertices in the theory must be of the Einstein--Hilbert type, as well as other constraints involving matter couplings to a scalar.  If an effective theory has vertices which do not satisfy these constraints, then there is superluminality at a scale parametrically the same as the scale suppressing the cubic couplings, which in effective theories of bi-gravity is generally the mass of the spin-2 particle.  Curing this superluminality   requires the presence of new physics at this scale.  Thus, in any UV-complete theory with these low energy degrees of freedom and a parametric gap to the next states, we can say the cubic vertices must satisfy our constraints.

In the purely massive case, we also have other IR constraints on UV completion coming from analytic dispersion relations~\cite{Cheung:2016yqr,Bonifacio:2016wcb,Bellazzini:2017fep,deRham:2017xox} applied to the four particle amplitude in the forward limit $t\rightarrow 0$.  When there are massless particles present these constraints become muddled due to $\sim 1/t$ singularities in the forward limit and possible violations of the Froissart bound, so it is difficult to extract any clean constraints.  Thus in the case of interest for this paper we do not have any such complementary constraints and the eikonal constraints are the only robust clues about UV completion that we are aware of.

When these constraints are applied to ghost-free bi-gravity, they reduce the possible two-parameter family of theories to the unique one-parameter sub-family which possesses a $\mathbb{Z}_2$ symmetry under the interchange of the two metrics.  Curiously, this is not the first time that a $\mathbb{Z}_2$-symmetric bi-gravity theory has been identified in the literature as possessing special properties beyond this symmetry.  In \cite{deRham:2012kf,Hassan:2012gz,deRham:2013wv} a $\mathbb{Z}_2$-symmetric theory was identified as the unique candidate theory for ``partially massless" gravity, an exotic representation of the de Sitter algebra that propagates fewer degrees of freedom than a usual massive spin-2 particle~\cite{Deser:1983tm}.  In this theory, the longitudinal mode of the massive spin-2 particle is entirely absent in the scalar-tensor sector of a certain high-energy limit (the ``decoupling limit").  This requirement fixes $c_3 = \frac{1}{4}$, just as we found above, and also $d_5 = -\frac{1}{32}$, where $d_5$ controls the quartic interactions of the theory. It would be interesting to extend our calculation to include next-to-leading-order terms in the eikonal limit, which would be sensitive to quartic vertices, and to check if $d_5 = -\frac{1}{32}$ is identified as a special point.

It is worthwhile to mention the assumptions that go into these constraints.  First of all, we assume that the eikonal approximation is in fact valid in the spin-2 case, {\it i.e.}, that summing over the ladder diagrams indeed captures the eikonal  limit of the full scattering amplitude.  The eikonal approximation for lower-spin particles in fact fails, as described in~\cite{Tiktopoulos:1971hi,Cheng:1987ga,Kabat:1992pz}.   A physical argument for its validity for spins $\geq 2$ and its failure for spins $<2$ is given in~\cite{Camanho:2014apa}, but it has not been proven at a technical level.
Secondly, we have assumed that the absence of asymptotic time advances is in fact a fundamental requirement of a UV complete theory.  As far as we are aware, there is no direct derivation of this requirement as a consequence of more fundamental $S$-matrix requirements like locality or analyticity.   It is possible that the presence of superluminality of this kind may not always lead to acausality, closed time-like curves, or consistency problems \cite{Babichev:2007dw,Geroch:2010da,Burrage:2011cr,Hassan:2017ugh}.  Conversely, even in the absence of superluminality of this kind, there could be other consistency issues that arise, possibly on other backgrounds.\footnote{According to \cite{Hassan:2017ugh}, problematic spacetimes admitting closed time-like curves are excluded in ghost-free bi-gravity provided that the square root matrix appearing in the bi-gravity action is specified in a proper way.} Finally, we have assumed that flat space is a solution out to length scales much larger than the Compton wavelength of the massive spin-2 particle.  In many cosmological applications of massive gravity or bi-gravity, this is typically not the case, since the horizon size is of order $m^{-1}$, and so the bounds derived do not directly apply.  It would be interesting to find analogous bounds directly in cosmological backgrounds.

There are a variety of situations where both massless and massive spin-2 states appear, and it would be interesting to explore how our constraints interact with these examples. It is already known that the way string theory is consistent with these eikonal constraints is rather intricate: it is not through having the cubic vertex combinations we have identified, but rather through cancellations between particles on the leading Regge trajectory~\cite{Camanho:2014apa,DAppollonio:2015fly}. Another oft-studied situation where massive spin-2 states arise coupled to gravity is in Kaluza--Klein reductions, where an infinite set of massive spin-2 particles appear. It would be illuminating to study the cubic couplings which arise in these cases.

Finally, we have focused on scattering amplitudes in flat space, but the AdS version of our constraints should connect to causality constraints on CFT correlation functions involving the stress tensor and a non-conserved spin-2 operator, with the gap to the next heaviest operator, $\Delta_{\rm gap}$, playing the same role as $m$ in our analysis. The analogous connection for pure stress tensor correlators has by now led to many interesting consequences, {\it e.g.},~\cite{Hofman:2008ar,Hofman:2009ug,Buchel:2009tt,Hartman:2016dxc,Afkhami-Jeddi:2016ntf}. Indeed, recently~\cite{Meltzer:2017rtf} have placed constraints on some mixed spin-2--stress tensor correlators using the Regge behavior of CFT correlators~\cite{Kulaxizi:2017ixa,Li:2017lmh,Costa:2017twz,Afkhami-Jeddi:2017rmx} and our results appear to be in qualitative agreement with theirs. It would be very interesting to further elucidate this connection.

\vspace{-.2cm}
\paragraph{Acknowledgements:} We would like to thank Clifford Cheung and Juan Maldacena for helpful conversations and correspondence.  RAR is supported by DOE grant DE-SC0011941 and Simons Foundation Award Number 555117.  AJ and RAR are supported by NASA grant NNX16AB27G.  

\renewcommand{\em}{}
\bibliographystyle{utphys}
\addcontentsline{toc}{section}{References}
\bibliography{bimetric-eikonal-JHEP2}

\end{document}